\documentclass[manuscript,screen]{acmart}
\AtBeginDocument{%
  \providecommand\BibTeX{{%
    \normalfont B\kern-0.5em{\scshape i\kern-0.25em b}\kern-0.8em\TeX}}}
    
\setcopyright{acmcopyright}
\copyrightyear{2024}
\acmYear{2024}

\usepackage{enumitem}
\usepackage{multirow}
\usepackage{colortbl}
\usepackage{url}
\usepackage{rotating}
\usepackage{xcolor}
\usepackage{soul}
\usepackage[ruled]{algorithm2e}
\usepackage{soul}
\usepackage[caption=false]{subfig}

\usepackage{booktabs}
\usepackage{nth}
\usepackage{threeparttable}
\usepackage{indentfirst}
\usepackage[most]{tcolorbox}
\usepackage[export]{adjustbox}
\usepackage[normalem]{ulem}
\usepackage{dashbox}

\newcommand{\specialcell}[2][c]{%
  \begin{tabular}[#1]{@{}c@{}}#2\end{tabular}}

\definecolor{custom-gray}{cmyk}{0, 0, 0, 0.7, 1.00}
\newtcbtheorem[no counter]{Summary}{\hskip-0.97em}{enhanced, drop shadow={black!50!white},
  coltitle=white,
  top=0.15in,
  attach boxed title to top left=
  {xshift=1.5em, yshift=-\tcboxedtitleheight/2},
  boxed title style={size=small, colback=custom-gray}
}{summary}

\begin{document}
\title{On the Model Update Strategies for Supervised Learning in AIOps Solutions}
\author{Yingzhe~Lyu}
\email{ylyu@cs.queensu.ca}
\affiliation{%
  \institution{Queen's University}
  \department{Software Analysis and Intelligence Lab (SAIL)}
  \city{Kingston}
  \state{ON}
  \country{Canada}
}

\author{Heng~Li}
\email{heng.li@polymtl.ca}
\affiliation{%
  \institution{Polytechnique Montreal}
  \department{Department of Computer and Software Engineering}
  \city{Montreal}
  \state{QC}
  \country{Canada}
}

\author{Zhen~Ming~(Jack)~Jiang}
\email{zmjiang@cse.yorku.ca}
\affiliation{%
  \institution{York University}
  \department{Department of Electrical Engineering \& Computer Science}
  \city{Toronto}
  \state{ON}
  \country{Canada}
}

\author{Ahmed~E.~Hassan}
\email{ahmed@cs.queensu.ca}
\affiliation{%
  \institution{Queen's University}
  \department{Software Analysis and Intelligence Lab (SAIL)}
  \city{Kingston}
  \state{ON}
  \country{Canada}
}

\begin{abstract}
  AIOps (Artificial Intelligence for IT Operations) solutions leverage the massive data produced during the operation of large-scale systems and machine learning models to assist software engineers in their system operations.
As operation data produced in the field are constantly evolving due to factors such as the changing operational environment and user base, the models in AIOps solutions need to be constantly maintained after deployment.
While prior works focus on innovative modeling techniques to improve the performance of AIOps models before releasing them into the field, when and how to update AIOps models remain an under-investigated topic.
In this work, we performed a case study on three large-scale public operation data: two trace datasets from the cloud computing platforms of Google and Alibaba and one disk stats dataset from the BackBlaze cloud storage data center.
We empirically assessed five different types of model update strategies for supervised learning regarding their performance, updating cost, and stability.
We observed that active model update strategies (e.g., periodical retraining, concept drift guided retraining, time-based model ensembles, and online learning) achieve better and more stable performance than a stationary model. 
Particularly, applying sophisticated model update strategies (e.g., concept drift detection, time-based ensembles, and online learning) could provide better performance, efficiency, and stability than simply retraining AIOps models periodically.
In addition, we observed that, although some update strategies (e.g., time-based ensemble and online learning) can save model training time, they significantly sacrifice model testing time, which could hinder their applications in AIOps solutions where the operation data arrive at high pace and volume and where immediate inferences are required.
Our findings highlight that practitioners should consider the evolution of operation data and actively maintain AIOps models over time. 
Our observations can also guide researchers and practitioners in investigating more efficient and effective model update strategies that fit in the context of AIOps.

\end{abstract}



\maketitle

\section{Introduction}
\label{sec:intro}

Large-scale software systems like Google and Amazon's cloud services are generating increasingly large volumes of operation data, such as alerting signals~\cite{chen2019outage}, events~\cite{elsayed2017learning}, logs~\cite{he2018identifying}, and resource usage metrics~\cite{xue2018spatial}.
On the one hand, such operation data provides rich information about the system's runtime behavior and health condition.
On the other hand, the soaring volume of operation data has become increasingly challenging for practitioners to collect, manage, analyze, and leverage. 
Therefore, AIOps~\cite{gartner2018aiops, dang2019aiops}, which stands for \underline{A}rtificial \underline{I}ntelligence for IT \underline{Op}eration\underline{s}, has been proposed to help practitioners address the challenges in the operations of large-scale software systems. 
AIOps solutions leverage machine learning (ML) techniques and operation data to support various goals in software and system operations, such as machine failure predictions~\cite{lin2018predicting, li2020predicting}, job failure predictions~\cite{elsayed2017learning, rosa2015catching}, disk failure predictions~\cite{xu2018improving, botezatu2016predicting, elsayed2017learning}, and service outage predictions~\cite{chen2019outage}.
Many of the proposed AIOps solutions have already shown promising benefits in practice. 
For example, Lin et al.~\cite{lin2018predicting} successfully applied their technique on one of Microsoft's large-scale cloud service systems to predict potential node failures based on historical data. 

Despite the breakthroughs in ML models and their applications in AIOps, many challenges are still associated with the maintenance and evolution of AIOps solutions following their deployment in the field. 
As operation data is usually produced in a dynamic environment where the hardware, software, and workloads can vary over time, the characteristics of the monitoring data are subject to constant changes~\cite{li2018adopting, lin2018predicting, li2020predicting, xu2018improving}.
As a result, AIOps models trained on data from the past may become outdated and perform poorly on the new data. Hence, these models must be constantly monitored and maintained to mitigate data evolution and preserve model performance.
Recently, software engineering tools such as continuous integration and DevOps have been applied to developing and maintaining ML models. For example, CD4ML~\cite{cd4ml} and MLOps~\cite{mlops} have been introduced under such contexts. 
However, no study systematically evaluates the model update strategies for AIOps models once deployed in the field.

Three characteristics of operation data combined make it unique from datasets in other application domains. 
First, operation data constantly evolves due to the dynamic operational environment (e.g., workload changes~\cite{li2018adopting}). Second, operation data is typically highly imbalanced (e.g., less than one in a thousand jobs fail~\cite{elsayed2017learning, rosa2015predicting}). Third, operation data usually contains a mixture of heterogeneous data types (e.g., temporal and spatial data~\cite{li2020predicting}). 
In this paper, we focus on the constant evolution of operation data and investigate the effectiveness of different model update strategies for supervised learning in AIOps solutions through a case study on three real-world operation datasets.
As different model update strategies can vary in performance and stability and add extra computational overhead to maintaining the performance of machine learning models, we focus our case study on the following three evaluation dimensions.

\begin{itemize}
    \item \textbf{Performance}. An AIOps solution should be accurate to provide actionable predictions~\cite{lin2018predicting,li2020predicting}.
    \item \textbf{Model updating cost}\footnote{In this work, we consider both the computational overhead (e.g., training time) and the cost-effectiveness of updating the model (i.e., the performance improvement achieved by each update of the model).}. As model updates can involve expensive computational efforts in training and verifying, a model update strategy with heavy maintenance overhead may not be desirable in AIOps solutions~\cite{dang2019aiops,li2020predicting}.
    \item \textbf{Stability}. For an AIOps solution to provide trustworthy predictions, its prediction performance should be stable~\cite{chen2019outage, lyu2021empirical, lyu2021interpretation}.
\end{itemize}

The \textit{performance} and \textit{stability} dimensions reflect the effects of model update strategies (i.e., how the update affects model performance), whereas the \textit{model updating cost} dimension captures the cost of reaching such effects (i.e., the computational effort needed to obtain the updated model). Together, the three dimensions measure the impact of adopting different model update strategies to mitigate the evolving nature of operation data.

In this work, we study the characteristics of data evolution in the AIOps context and evaluate different model update strategies for maintaining AIOps solutions in terms of their performance, updating cost, and stability. 
As the properties of operation data can shift significantly in different application scenarios, we further limit our scope to automated prediction of failures with supervised learning algorithms, which is a practical and popular type of problem faced by industry~\cite{he2018identifying, ding2012healing, xue2016managing}.
Specifically, we perform a case study on three publicly available operation datasets in this scope: the Google cluster trace dataset~\cite{clusterdata}, the Backblaze disk statistics dataset~\cite{backblaze}, and the Alibaba GPU cluster trace data~\cite{AlibabaCluster}.
Both the Google and Backblaze datasets are studied extensively in prior AIOps studies (e.g., ~\cite{elsayed2017learning,mahdisoltani2017proactive,rosa2015catching,rosa2015predicting,botezatu2016predicting, lyu2021empirical}) and the Alibaba dataset is relatively new with no work of failure prediction has been done to our best knowledge.
In particular, we focus on evaluating various model update strategies for AIOps tasks that predict job failures~\cite{elsayed2017learning} on the Google and Alibaba cluster trace datasets and disk failures~\cite{botezatu2016predicting, mahdisoltani2017proactive} on the Backblaze disk stats dataset.

The contributions of this paper are:
\begin{enumerate}
    \item This is the first work that evaluates model update strategies in the context of AIOps. As operation data is unique in characteristics from datasets in other domains due to its constant evolution, extreme data imbalance, and heterogeneous data types, our findings can provide specified insights in the AIOps context.
    \item Our case study shows that the characteristics of operation data could change drastically over time, which suggests that practitioners should carefully consider such data evolution in their modeling decisions.
    \item Our findings suggest that the emerging new strategies (e.g., online learning and time-based ensemble) are capable of generating better performance compared to the traditional retraining strategy, although these strategies may still need improvement to be used in practice (discussed in Section~\ref{sec:discussion}).
    \item Our findings suggest that, instead of only considering retraining their AIOps models periodically, AIOps practitioners could consider more sophisticated strategies, such as applying concept drift detection to determine when to maintain their models or using time-based ensemble strategies to update models.
    \item The approaches and findings presented in this paper can provide insights for future research that investigates more efficient and effective model update strategies that fit in the context of AIOps (e.g., more efficient ways of detecting concept drifts and more efficient time-based ensemble strategies).
    \item We share a replication package which includes our code for preprocessing the studied operation datasets and constructing the models\footnote{\url{https://github.com/SAILResearch/suppmaterial-20-yingzhe-AIOpsEvolvability}).}, so that others in the research community can replicate or extend our work.
\end{enumerate}

\noindent\emph{Paper organization.}
The rest of the paper is organized as follows: Section~\ref{sec:background} provides background information and discusses related works.
Section~\ref{sec:subjects-pre-study} describes our studied datasets, data preparation process, and preliminary study on the datasets. 
Sections~\ref{sec:experimental_design} presents our experiment design, and Section~\ref{sec:results} presents our experiment results.
Section~\ref{sec:discussion} provides further discussions of our experiment results.
Section~\ref{sec:threats} discusses the threats to the validity of our findings, and Section~\ref{sec:conclusion} concludes our paper.

\section{Related Work}
\label{sec:background}

This paper aims to evaluate different model update strategies for supervised learning in AIOps solutions.
In this section, we first present prior studies on AIOps solutions (e.g., ~\cite{lin2018predicting, li2020predicting, elsayed2017learning, rosa2015catching, botezatu2016predicting, mahdisoltani2017proactive, xu2018improving, chen2019outage}) and their practice of dealing with data evolution. 
We then discuss prior works in general areas that deal with data evolution for ML models, which may be applicable in the AIOps context to support the maintenance of AIOps solutions.

\subsection{Prior research on AIOps solutions}

Prior works proposed various AIOps solutions for addressing different problems in the operations of large-scale software and systems, including incident prediction~\cite{lin2018predicting, li2020predicting, elsayed2017learning, rosa2015catching, botezatu2016predicting, mahdisoltani2017proactive, xu2018improving, chen2019outage}, anomaly detection~\cite{he2018identifying, lim2014identifying}, ticket management~\cite{xue2018spatial, xue2016managing}, issue diagnosis~\cite{luo2014correlating}, and self healing~\cite{ding2012healing, ding2014mining, lou2013software, lou2017experience}.
For example, Lin et al.~\cite{lin2018predicting} and Li et al.~\cite{li2020predicting} leverage temporal data (e.g., CPU and memory utilization metrics, alerts), spatial data (e.g., node locations), and configuration data (e.g., memory size) to predict node failures on large-scale cloud computing platforms.
El-Sayed et al.~\cite{elsayed2017learning} and Rosa et al.~\cite{rosa2015catching} predict job failures in the Google cloud computing platform using trace data on the cloud servers.
Botezatu et al.~\cite{botezatu2016predicting}, Mahdisoltani et al.~\cite{mahdisoltani2017proactive}, and Xu et al.~\cite{xu2018improving} leverage monitoring data to predict disk failures in the operations of large-scale cloud platforms.

Prior works generally use the following two strategies to maintain the models in AIOps solutions: 
1) use stationary models~\cite{elsayed2017learning, rosa2015predicting, botezatu2016predicting, mahdisoltani2017proactive, chen2019outage, xue2018spatial} (a.k.a., no model maintenance) or 2) periodically retrain models~\cite{lin2018predicting, li2020predicting}.
Using a stationary model may lead to an outdated model that suffers from performance degradation and impacts user experience, while periodically retraining the model could be very expensive (e.g., resources and human efforts involved in updating, verifying, and integrating the model)~\cite{li2020predicting}.
Therefore, in this work, we study various strategies for model updating and their impact on the performance, model updating cost, and stability of AIOps solutions.

\subsection{Prior research on dealing with data evolution}
Although prior works in AIOps usually use stationary models or periodically retraining strategies, studies in other areas have proposed various approaches to deal with data evolution and maintain ML models.
For example, prior works in the data mining area propose various methods for detecting concept drift~\cite{gama2004learning, harel2014concept, nishida2007detect, wang2013concept}, so that the models could be updated timely and more effectively compared with the periodical retraining strategy.
Prior studies also propose online models (e.g., Hoeffding Trees~\cite{domingos2000mining}) for handling concept drift by enabling the models to learn incrementally.
In addition, time-based ensemble methods are proposed to handle concept drift by building individual classifiers on data chunks from different time periods, then aggregating the models through a voting mechanism~\cite{dongre2014review}.

\subsubsection{Concept drift detection and mitigation}
In machine learning and data mining, the distribution of data and the relationship between the variables may evolve over time, known as concept drift~\cite{wang2010mining, wang2003mining, tsymbal2004problem, nishida2007detect}. 
Concept drift may negatively impact the performance of a model trained on previous data as the data evolves.
Prior works propose algorithms for detecting concept drift to mitigate the impact of concept drift~\cite{gama2004learning, harel2014concept, nishida2007detect}.
For example, Gama et al.~\cite{gama2004learning} propose the Drift Detection Method (DDM) that detects the concept changes by tracking the variation of prediction error, assuming that a significant increase in the testing error suggests the change of the concept.
Harel et al.~\cite{harel2014concept} propose another method, namely PERM, to detect concept drift with the assumption that random permutation of examples from the same distribution should not change the model performance.
Nishida et al.~\cite{nishida2007detect} propose the STEPD concept drift detection method using a statistical test of equal proportions.
It assumes that the prediction accuracy on data from a recent time window should be equal to the overall prediction accuracy if the target concept is stationary. A significant decrease in the prediction accuracy then suggests concept drift.
Once concept drift is detected, the current model is usually updated or retrained using the newest data~\cite{gama2004learning, harel2014concept, nishida2007detect, gama2014survey}.
In this paper, we name the model update strategy of retraining model after detection of concept drift as concept drift guided retraining strategy (See Section~\ref{sec:experimental_design} for more detail)

\subsubsection{Time-based ensemble models}
Prior research also proposed time-based ensemble models to handle concept drift~\cite{street2001streaming, wang2003mining, brzezinski2014reacting, dongre2014review, cano2020kappa, hoens2012learning, gama2014survey, minku2011ddd, minku2009impact}.
Time-based ensembles combine individual base classifiers trained from smaller time periods of the data.
For example, Street and Kim propose the Streaming Ensemble Algorithm (SEA)~\cite{street2001streaming}, a majority-voting ensemble approach that constantly replaces the weakest base classifier in the ensemble. 
It builds a new classifier on the most up-to-date data chunk and decides whether to accept it into the ensemble using a quality measure tested on the following data chunk that considers both the accuracy and diversity of base classifiers in the ensemble.
Wang et al. propose Accuracy Weighted Ensemble (AWE)~\cite{wang2003mining}, another time-based ensemble model that uses weighted voting rather than the simple majority voting as used in SEA. AWE calculates a weight for each classifier in the ensemble based on its error rate on the newest data.

\subsubsection{Online learning models}
Online learning algorithms provide a method for incrementally learning from a large volume of data samples.
Domingos and Hulten~\cite{domingos2000mining} propose the Very Fast Decision Tree (VFDT or Hoefdding Tree) for learning from data streams by updating the tree on the fly. 
The authors claim that, given sufficient samples, Hoefdding Tree can produce nearly identical results as a conventional batch tree learner.
Gomes et al.~\cite{gomes2017adaptive} propose the Adaptive Random Forest (ARF), a Random Forest algorithm for handling concept drift in evolving data streams.
The ARF algorithm incorporates an online bagging algorithm to learn incrementally from streaming data. It also applies strategies, including drift detection and weighted voting, to mitigate the concept drift problem.
There are also time-based ensembles utilizing online learning models as base classifiers (e.g., ADE~\cite{dongre2014review}, AUE~\cite{brzezinski2014reacting}, KUE~\cite{cano2020kappa}). 
For example, Brzezinski et al. propose the Accuracy Updated Ensemble (AUE) model~\cite{brzezinski2014reacting}, which uses online base classifiers and a weighted voting mechanism to deal with concept drift.
Cano and Krawczyk propose Kappa Updated Ensemble (KUE)~\cite{cano2020kappa}, a combination of online and block-based ensemble approaches that use the Kappa statistic for dynamic weighing and selection of base classifiers.

\hfill \break
In this work, we study the impact of the model update strategies (i.e., periodical retraining, concept drift guided retraining, time-based ensembles, and online learning strategies) on AIOps models. 
Specifically, we apply various model update strategies for prediction tasks on the studied operation datasets and evaluate their impact on the three dimensions of performance, model updating cost, and stability.

\section{Case Study Subjects and Preliminary Study} \label{sec:subjects-pre-study}
In this section, we first describe our studied datasets, then perform a preliminary study on the datasets.

\subsection{Case Study Subjects}
\label{sec:case_study}

In order to evaluate different strategies for model updating in the context of AIOps, we perform a case study on three large-scale operation datasets: the Google cluster trace dataset~\cite{clusterdata}, the Backblaze disk stats dataset~\cite{backblaze}, and the Alibaba GPU cluster trace data~\cite{AlibabaCluster}.
We choose to study these three datasets because: 1) they are publicly available; 2) they are large-scale datasets and cover relatively long operation periods (i.e., months to years), which enables us to examine model update strategies over the evolution of the data.
In addition, prior works have widely studied the first two datasets in particular for predicting job failures on the Google dataset~\cite{elsayed2017learning, rosa2015predicting} and predicting disk failures on the Backblaze dataset~\cite{botezatu2016predicting, xu2018improving, mahdisoltani2017proactive}.
In this work, we focus on predicting job outcomes (i.e., failure or not) on the Google and Alibaba cluster trace datasets and predicting disk failures on the Backblaze disk stats dataset.

\subsubsection{Google Cluster Trace Dataset}

\begin{figure}[!htbp]
    \centering
    \subfloat[Google data schema.]{
        \dbox{\includegraphics[width=0.296\textwidth, valign=t]{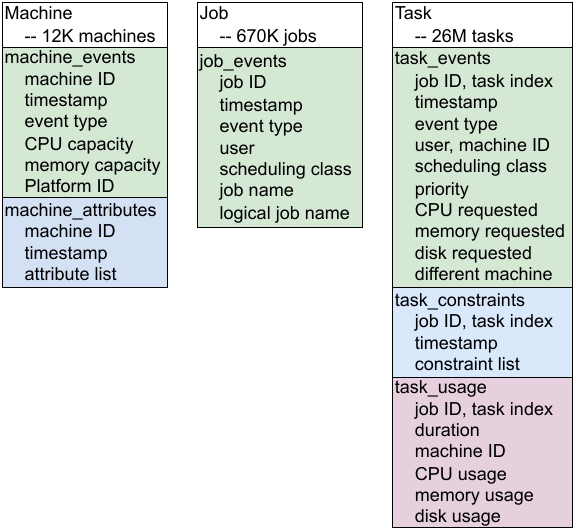}}
        \label{fig:google_schema}
    }
    \subfloat[Backblaze data schema.]{
        \dbox{\includegraphics[width=0.27\textwidth, valign=t]{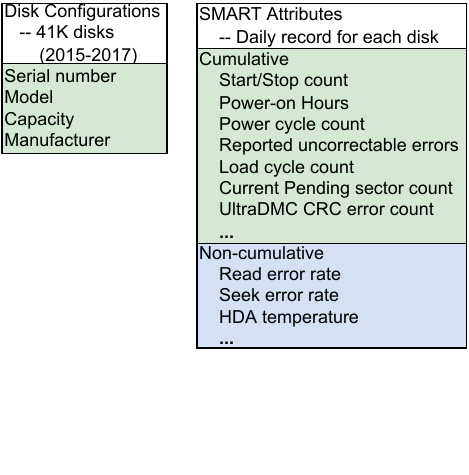}
        }
    \label{fig:disk_schema}
    }
    \subfloat[Alibaba data schema.]{
        \dbox{\includegraphics[width=0.334\textwidth, valign=t]{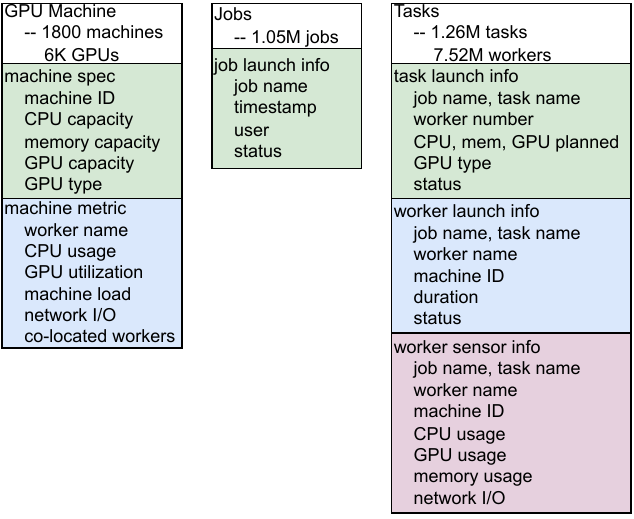}
        }
    \label{fig:alibaba_schema}
    }
    \caption{Data schema for our studied datasets. 
    Each colored box represents a data table: a line of the table name followed by lines describing the data fields.
    For the Google and Alibaba datasets, each table (e.g., machine\_events) is one or multiple CSV files containing the fields described in the box. 
    For the Backblaze dataset, the tables represent the logical view, while the physical data is stored as daily snapshots of each disk's attributes.
    }
    \label{fig:schema}
\end{figure}

The cluster data released by Google in 2011 contains the trace data of a production cluster with about 12K machines in 29 days for 670K jobs and 26M tasks~\cite{chen2014failure}. 
The data features workload arrives at a cell (i.e., a set of machines that share a common cluster-management system) in the form of jobs. Each job comprises one or more tasks, each of which is scheduled on a single machine.
Figure \ref{fig:google_schema} shows the dataset schema and information provided in the Google cluster trace dataset.

Following prior works~\cite{elsayed2017learning, rosa2015predicting}, our goal on the Google cluster data is to predict whether a job will fail or not (i.e., terminated for any reason before successfully completed) using the information at job submission and the monitoring data in the first five minutes of the job execution. 
In the Google cluster trace data, each job has several events, and each is associated with a transit (e.g., submit, schedule, evict, fail, kill, finish, lost, update) among the states (e.g., unsubmitted, pending, running, dead) in the job's life cycle. 
We consider a job fails if its final state is ``fail'', same as in prior works~\cite{elsayed2017learning, rosa2015predicting}.

Similarly to El-Sayed et al.~\cite{elsayed2017learning}, we predict the job failures using the configuration and temporal features.
Configuration features are values determined upon job submission, such as the requested CPU, memory, and disk space.
In contrast, temporal features are values that change during a job's execution, such as the mean and standard variation of CPU, memory, and disk space usage by a job over the first 5 minutes since job submission.
The detailed list of the features for job failure prediction is described in Table~\ref{tab:features_google}, where the first 9 are configuration features while the latter 6 are temporal features. 

We remove the jobs that are not completed or whose records are lost during execution, as the final states of these jobs are missing.
We further remove the jobs that start from the last day (i.e., the 29th day), as these jobs are more likely to fail than complete before the data cutoff time (the completed jobs typically last longer than the failed ones), which may cause data collection bias on the distribution of failed and completed jobs. In fact, we observed a much higher job failure rate from the jobs starting on the last day.   
In addition, we removed jobs that finished in less than five minutes since their submission as they have not generated sufficient metrics for prediction. 
We also observe that a large proportion of these jobs are failed or terminated right after submission, thus they do not cause significant overhead to the computing resources.
In the end, we successfully extracted 627K (out of 670K) job samples from the first 28 days' trace data. 

\subsubsection{Backblaze Disk Stats Dataset}

The Backblaze dataset describes the statistics of the hard drives in the Backblaze data center~\cite{backblaze}.
The dataset contains daily snapshots of operational hard drives in the data center, including drive information (e.g., model, disk capacity) and SMART (Self-Monitoring, Analysis, and Reporting Technology) statistics, where SMART is a manufacturer-implemented system for the monitoring and early detection of errors. 
Figure \ref{fig:disk_schema} shows the dataset schema and information provided in the Backblaze disk stats dataset.
The Backblaze disk stats dataset contains hard drive monitoring data collected from 2013 to 2020. 
Initially, from 2013 to 2014, the trace captured 40 different SMART attributes; 
then, from 2015 to 2017, there are 45 SMART attributes; 
starting from the fourth quarter of 2018, 62 different SMART attributes are in the data.
Despite the change of monitored attributes, the data type and the monitoring interval (i.e., daily) are kept consistent from 2013 to 2020.
We focus on the data collected from 2015 to 2017 because: 1) the subset contains a large number of samples (i.e., over 40M samples), and 2) the subset contains a fixed set of SMART attributes while the data in other periods contains less (before 2015) or more (after 2017) SMART attributes.

Our goal on the Backblaze dataset is to predict hard drive failures (i.e., sector error) within a given future time period (i.e., one week) based on the monitoring data captured during a period of time (i.e., one week) in the past, similar to the prior works~\cite{mahdisoltani2017proactive, botezatu2016predicting}. 
We consider a disk fails if its ``sector error count'' SMART attribute increases (i.e., observe sector errors) in the given future time period, same as described in the prior work~\cite{mahdisoltani2017proactive}.
The SMART attributes that exist in the Backblaze dataset can be categorized into two types: cumulative attributes whose values are accumulated counts over the disk's lifetime, such as the ``reallocated sectors count''; and noncumulative attributes whose values reflect only the current status, such as the ``read error rate''. 
Knowing the recent changes in cumulative attributes rather than their raw values might be more insightful.
Therefore, we capture both the value change in the past time period and the raw value in the last day of the one-week past window as features for cumulative attributes while only capturing the last day's value for noncumulative attributes.
As a result, we collect 11 features from the raw values and 8 features derived from the raw values' differences.
The detailed list of our used features for disk failure prediction is described in Table~\ref{tab:features_backblaze}.
The collected features are the same as those used in prior work~\cite{mahdisoltani2017proactive}, the most predictive ones selected from all the traced SMART attributes; all 19 features are temporal features.
We then collect data samples along the sliding one-week time window and only track the disks that are alive during the whole time window.
As a result, we extract 41M samples from the daily snapshots between 2015 and 2017.

\begin{table}[htbp]
    \centering
    \caption{Features for Google job failure prediction.}
    \label{tab:features_google}
    \begin{tabular}{p{14em}p{31em}}
        \toprule
        Feature& Description\\
        \midrule
        User ID& Unique ID of the user who submits the job.\\
        Logical Job Name& A hashed job identifier contains information from several internal name fields. Different executions of the same program usually have the same logical name.\\
        Scheduling class& A nominal field represented by a single number (e.g., 3 representing a more latency-sensitive task and 0 representing a nonproduction task) that affects machine-local policy for resource access.\\
        Num Tasks& The number of tasks contained in a job.\\
        Priority& A nominal field indicates how effectively it can utilize system resources.\\
        Different machine& A flag that indicates tasks must be scheduled to execute on a different machine than any other currently running tasks in the job.\\
        Requested CPU/Memory/Disk& Requested CPU/memory/disk resources by tasks in a job.\\
        Mean CPU/Memory/Disk usage& Mean CPU/memory/disk usage over 5 minutes since job submission.\\
        Sd CPU/Memory/Disk usage& Standard variation of the CPU/memory/disk usage over 5 minutes since job submission.\\
        \bottomrule
    \end{tabular}
\end{table}

\begin{table}[htbp]
    \centering
    \caption{Features for Backblaze disk failure prediction.}
    \label{tab:features_backblaze}
    \begin{threeparttable}
    \begin{tabular}{p{18em}p{27em}}
    \toprule
    Feature& Description\\
    \midrule
    Read Error Rate (Non-cumulative)\tnote{1}& Frequency of errors while reading raw data from a disk.\\
    Start/Stop Count (Cumulative)& Number of spindle start/stop cycles.\\
    Reallocated Sectors Count (Cumulative)& Quantity of remapped sectors.\\
    Seek Error Rate (Non-cumulative)& Frequency of errors while positioning.\\
    Power-On Hours (Cumulative)& Number of hours elapsed in the power-on state.\\
    Power Cycle Count (Cumulative)& Number of power-on events.\\
    Reported Uncorrectable Errors (Cumulative)& Number of reported uncorrectable errors.\\
    Load Cycle Count (Cumulative)& Number of cycles into landing zone position.\\
    HDA Temperature (Non-cumulative)& Temperature of a hard disk assembly.\\
    Current Pending Sector Count (Cumulative)& Number of unstable sectors (waiting for remapping).\\
    UltraDMC CRC Error Count (Cumulative)& Number of CRC errors during UDMA mode.\\
    \bottomrule
    \end{tabular}
    \begin{tablenotes}
    \item[1] For cumulative SMART attributes, both their raw value of the last day and the difference during the training period are extracted as features, while for non-cumulative attributes we use only the raw value of the last day as feature.
    \end{tablenotes}
    \end{threeparttable}
\end{table}

\subsubsection{Alibaba GPU Cluster Trace Dataset}

The GPU cluster trace data from Alibaba provides traces of workloads collected from the operation of a large-scale data center~\cite{AlibabaCluster}.
The trace data is collected from runtime information on over 6,500 GPUs across about 1,800 machines in a period of 2 months spanning from July to August of 2020~\cite{weng2022mlaas}.
The data features ML jobs submitted by various users. 
Once a user submits a job, the job is translated into multiple tasks of different roles. 
Subsequently, each task is then allocated to machines in the format of instances. 
Figure~\ref{fig:alibaba_schema} illustrates the trace schema and available information provided in the Alibaba trace dataset.
Similar to the Google cluster dataset, sensitive fields like username and job name are desensitized to protect users' privacy.

The dataset is relatively new (released in 2021), and we have not found prior works that conduct prediction tasks on this dataset to the best of our efforts.
Therefore, we define the task as predicting job outcomes using the information available in the first five minutes since job submission, similar to our case study on the Google cluster dataset. 
We then define and extract 12 predictive features on our own and list the details in Table~\ref{tab:features_alibaba}, where the first $6$ features are configurational while the latter $6$ are temporal.
The dataset contains cluster monitoring data for a total of 69 days (around nine weeks). 
To avoid abnormality (e.g., truncated and untracked jobs) on data samples close to the beginning and end of the trace data, we initiate feature extraction from the fourth day since the trace starts and collect features for a total of 8 weeks.
Similar to our handling method on the Google cluster dataset, we also removed unfinished jobs and jobs ending in less than five minutes and extracted 701K out of the total of 1.26M jobs.

\begin{table}[htbp]
    \centering
    \caption{Features for Alibaba job failure prediction.}
    \label{tab:features_alibaba}
    \begin{tabular}{p{11em}p{32em}}
        \toprule
        Feature& Description\\
        \midrule
        User ID& Unique ID of the user who submits the job.\\
        Num Tasks& Number of the unique task names (e.g., ps, worker, evaluator) in a job.\\ 
        Num Instances& Number of the instances contained in a job.\\
        Planned CPU& Number of CPU cores requested in percentage. \\
        Planned Memory& Sum of GBs of main memory requested by tasks in a job. \\
        Planned GPU& Sum of the number of GPUs requested by tasks in a job. \\
        Mean CPU Usage& Number of CPU cores used in percentage.\\
        Mean GPU utilization& Number of GPUs used in percentage.\\
        Mean memory usage& Mean usage of main memory in the first 5 minutes since job submission.\\
        Max memory usage& Maximum amount of memory used over 5 minutes since job submission.\\
        Mean GPU memory usage& Mean usage of GPU memory over 5 minutes since job submission.\\
        Max GPU memory usage& Maximum amount of GPU memory used over 5 minutes since job submission.\\
        \bottomrule
    \end{tabular}
\end{table}

\subsection{Preliminary Study}
\label{sec:preliminary}

As the operational environment and the workloads of large-scale systems are constantly evolving~\cite{li2018adopting}, the performance of AIOps models may be impacted by the evolution of the operation data~\cite{li2020predicting, lin2018predicting}.
In order to understand the need for maintaining AIOps models, we investigate the data evolution in our studied operation datasets.

\subsubsection{Approach}

We study the evolution of the operation data along two dimensions: 1) the evolution of data volume and 2) the evolution of data distribution.
As described in Section~\ref{sec:case_study}, the schema (i.e., monitored attributes) of the Backblaze dataset also changes over time. 
We do not consider such schema change in this work but leave it to future work.

\noindent\textbf{Evolution of the volume of the data}.
We first partition the data into multiple time periods according to their timestamps. 
For the Google cluster trace data, we partition the entire 28-day trace data into 28 one-day time periods; 
for the Backblaze disk stats data, we partition the entire 3-year monitoring data into 36 one-month time periods;
for the Alibaba GPU cluster data, we partition the 2-month GPU trace data into 8 one-week time periods.
We choose the time window sizes since prior studies have applied similar time periods in updating their AIOps models.
For example, Lin et al.~\cite{lin2018predicting} update their model deployed in a production cloud service system with data from the most recent one-month period.
Similarly, Li et al.~\cite{li2020predicting} consider retraining their model periodically, and they also apply a one-month window.
Besides, Xu et al.~\cite{xu2018improving} perform a daily model update with the data in a 90-day sliding window.
After partitioning the data into time periods, we analyze how the number of samples (i.e., data volume) changes across different time periods.

\noindent\textbf{Evolution of the distribution of the data}.
Similarly, we partition the data into multiple time periods and analyze how the distribution of the variables evolves across different time periods in the studied datasets.
For the dependent variable, we measure its distribution (i.e., the proportion of failed jobs or disks) in each time period.
We also estimate the statistical difference between the distributions of the dependent variable in different time periods using a two-proportion Z-test~\cite{fisher2011testing}, which compares whether the job or disk failure rates in two time periods are statistically different.
In addition, we use Cliff's delta~\cite{macbeth2011cliff} effect size to calculate the magnitude of the difference between the distribution of the dependent variable in different time periods.
Since our dependent variables (i.e., job failure and disk failure) are binary, we first calculate the log odds ratio between the distributions of the dependent variable in two time periods, then convert it to Cliff's delta effect size~\cite{cooper2019handbook}.
We apply the thresholds provided in Romano et al.~\cite{romano2006appropriate} to assess the magnitude of Cliff's $\delta$: Negligible, $\delta < 0.147$; Small: $\delta < 0.33$; Medium: $\delta < 0.474$; Large: $\delta \ge 0.474$.

\subsubsection{Results}
\begin{figure}
    \centering
    \subfloat[Google]{\label{fig:data_size_google}{
    \includegraphics[width=0.3\textwidth]{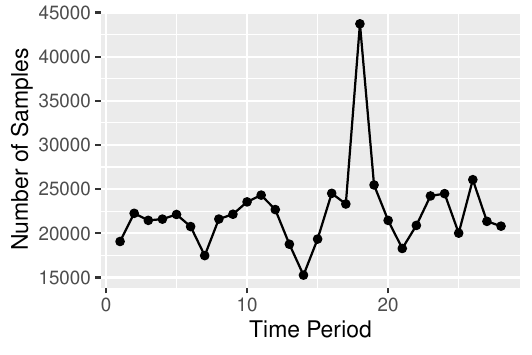}}}\hfill
    \subfloat[Backblaze]{\label{fig:data_size_backblaze}{
    \includegraphics[width=0.3\textwidth]{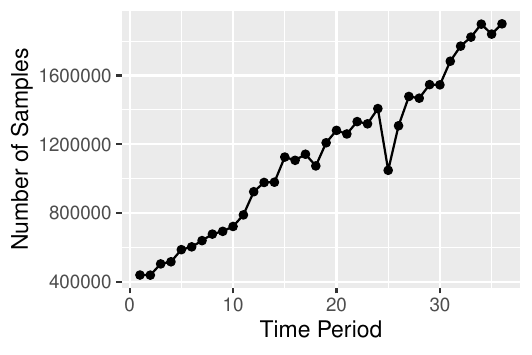}}}\hfill
    \subfloat[Alibaba]{\label{fig:data_size_alibaba}{
    \includegraphics[width=0.3\textwidth]{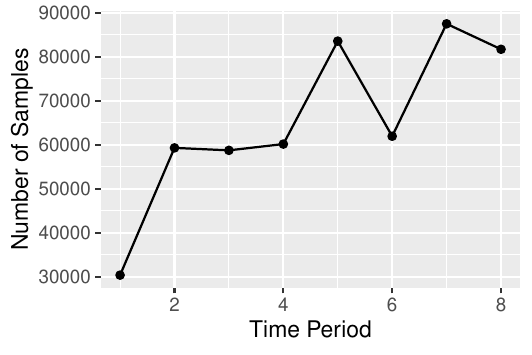}}}
    \caption{Number of samples in different time periods of the studied datasets.}
    \label{fig:data_sizes}
\end{figure}

\begin{figure}
    \centering
    \subfloat[Google]{\label{fig:failure_rate_google}{
    \includegraphics[width=0.3\textwidth]{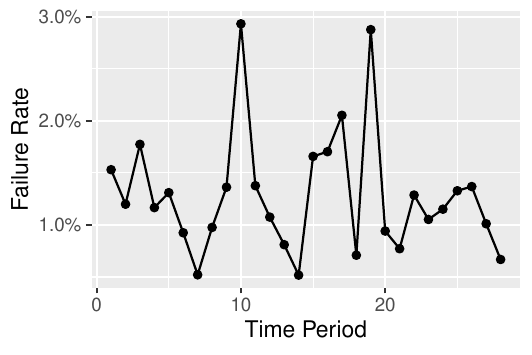}}}\hfill
    \subfloat[Backblaze]{\label{fig:failure_rate_backblaze}{
    \includegraphics[width=0.3\textwidth]{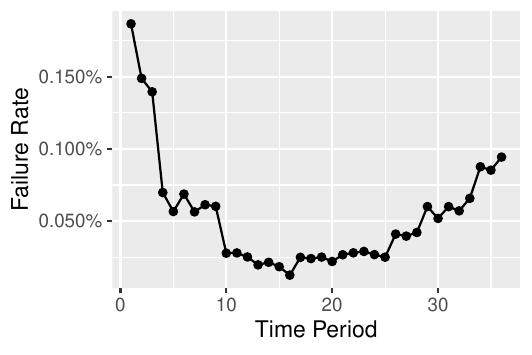}}}\hfill
    \subfloat[Alibaba]{\label{fig:failure_rate_alibaba}{
    \includegraphics[width=0.3\textwidth]{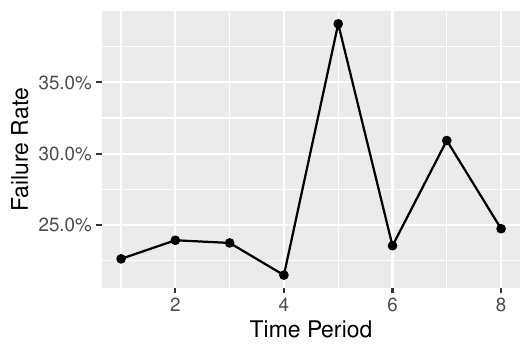}}}
    \caption{Failure rates in different time periods of the studied datasets.}
    \label{fig:failure_rates}
\end{figure}

\begin{figure}
    \centering
    \subfloat[Google]{\label{fig:corr_dep_google}{
    \includegraphics[width=0.3\textwidth]{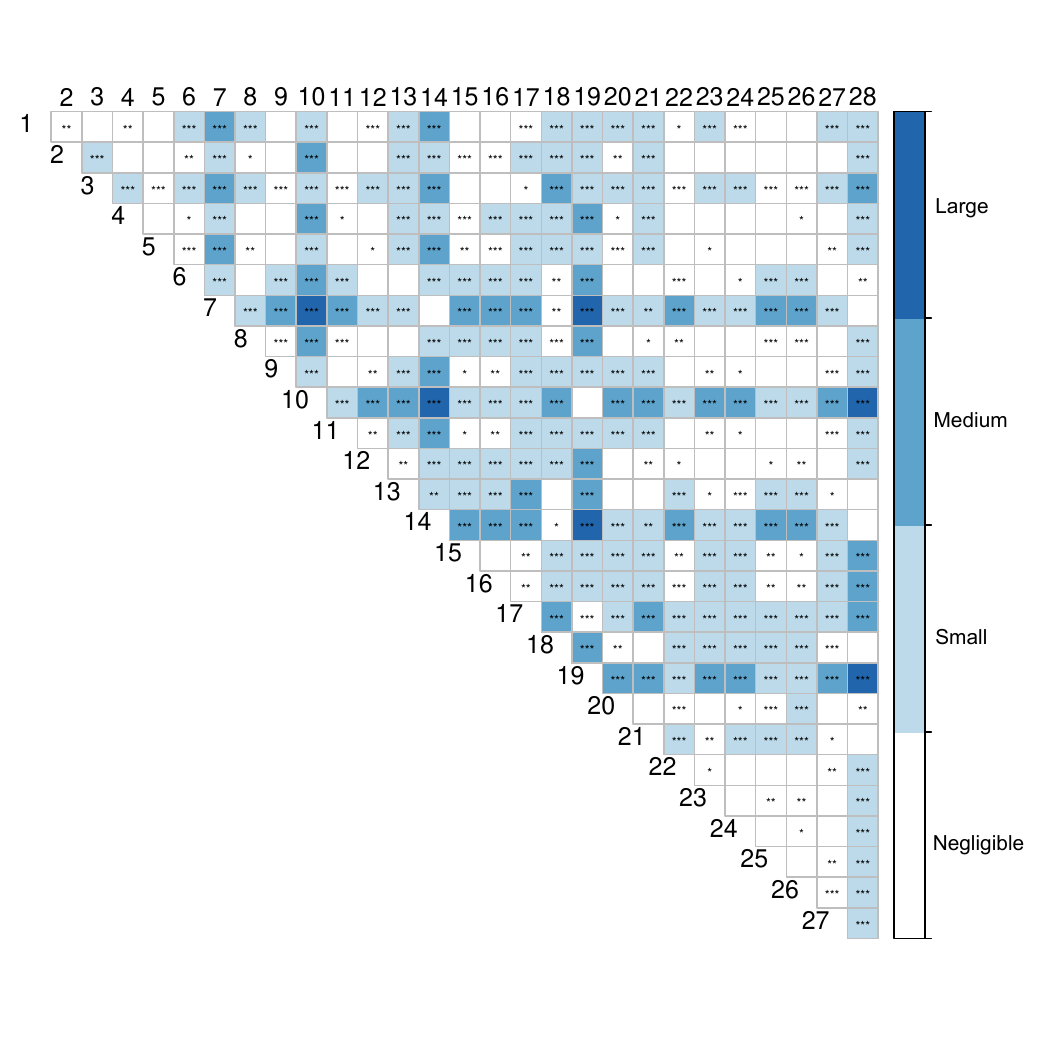}}}
    \subfloat[Backblaze]{\label{fig:corr_dep_backblaze}{
    \includegraphics[width=0.3\textwidth]{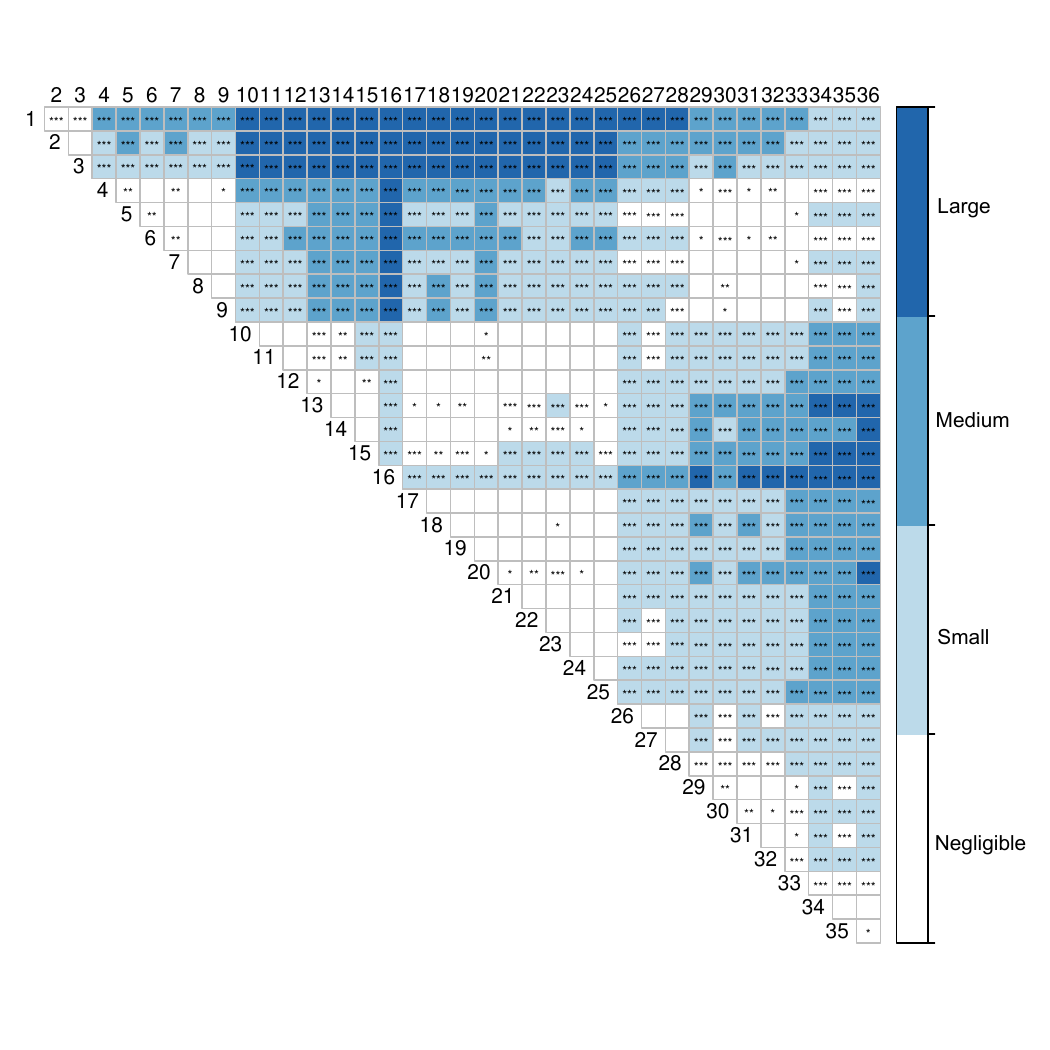}}}
    \subfloat[Alibaba]{\label{fig:corr_dep_alibaba}{
    \includegraphics[width=0.3\textwidth]{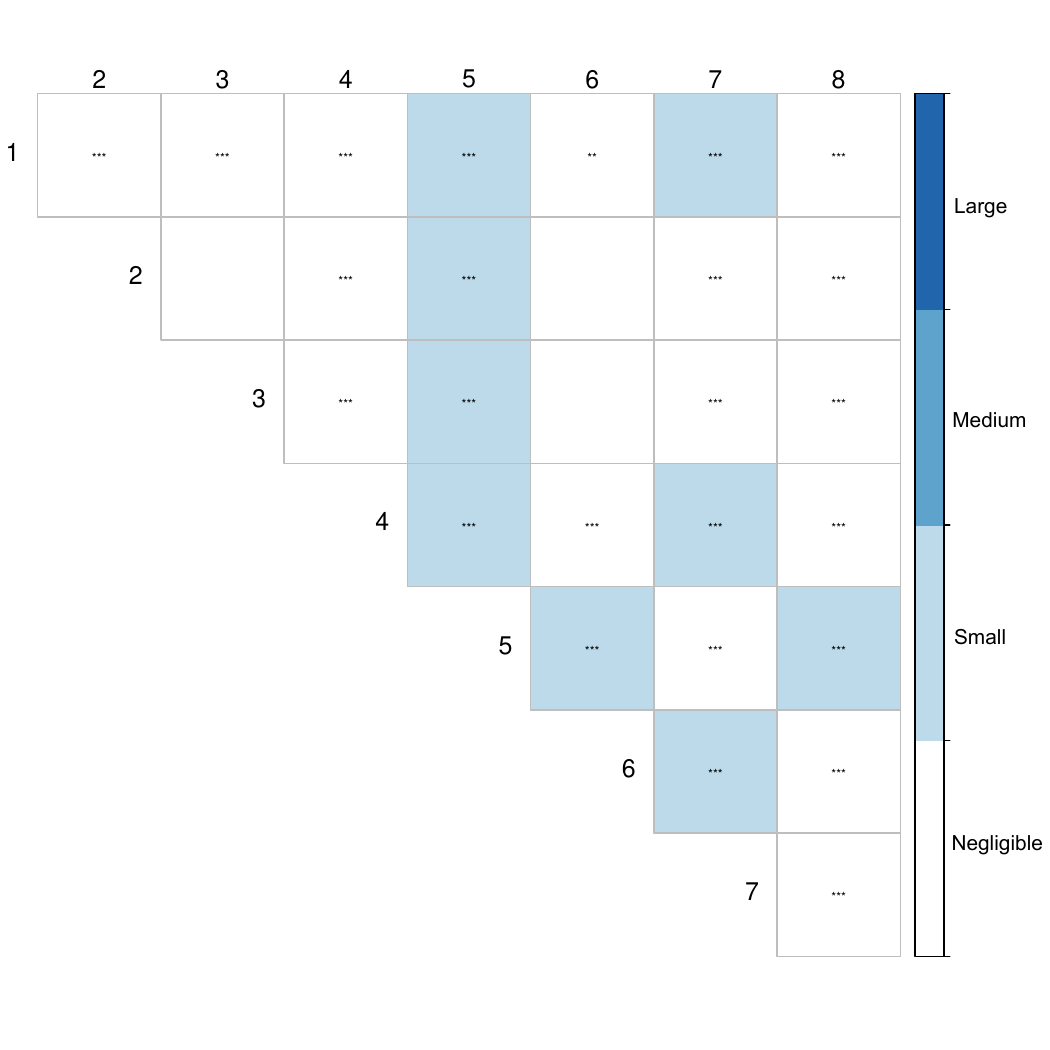}}}
    \caption{Statistical difference of dependent variables in different time periods of the studied datasets. The symbols in each cell indicate the statistical significance of the failure rate difference: (blank) $p\ge 0.05$; * $p < 0.05$; ** $p<0.01$; *** $p<0.001$. The color indicates the effect size of the failure rate difference using $\delta$: Negligible, $\delta < 0.147$; Small: $\delta < 0.33$; Medium: $\delta < 0.474$; Large: $\delta \ge 0.474$.}
    \label{fig:corr_dep}
\end{figure}

Figure~\ref{fig:data_sizes} shows the change in the number of samples across different time periods.
Similarly, Figure~\ref{fig:failure_rates} shows the distributions of the dependent variables (i.e., the daily Google job failure rate and the monthly Backblaze disk failure rate) across different time periods.
Figure~\ref{fig:corr_dep} further shows the statistical difference of dependent variable distributions in different time periods.

\textbf{The volume and distribution of the operation data constantly evolve over time.}
For the Google cluster trace data, as shown in Figure~\ref{fig:data_size_google}, the number of samples in each day fluctuates over time. 
For example, the number of samples on day 18 surges to over 43K from the average value of 22K. 
For the Backblaze dataset, as shown in Figure~\ref{fig:data_size_backblaze}, the number of samples in each month constantly increases, except for the 18th month (June 2016), where the number of samples drops over 25\% from the previous time period.
For the Alibaba GPU cluster trace data, as shown in Figure~\ref{fig:data_size_alibaba}, the number of samples climbs up from 30K in the first week to 59K in the second week, then fluctuates between 58K and 88K in the following weeks.

As shown in Figure~\ref{fig:failure_rates}, the distribution of dependent variables on the two datasets also changes over time.
The job failure rate fluctuates between 0.5\% and 3\% on the Google dataset and between 21.5\% to 39.1\% on the Alibaba dataset, while the Backblaze disk failure rate decreases from 0.19\% in the first month to 0.01\% (i.e., over 90\% decrease) in the $16$th month and gradually rises to around 0.09\% in the last month. 
Our statistical analysis (Figure~\ref{fig:corr_dep}) shows that the difference between the distributions of the dependent variables in different time periods can be statistically significant. 
We observe that the difference in the distributions is significant (i.e., p < 0.05) between most pairs of time periods.
We also notice the distributions of the dependent variable can show \texttt{medium} to \texttt{large} differences in terms of effect sizes between two time periods.


\textbf{Drastic changes in data distribution could also happen in adjacent time periods}.
For the Backblaze dataset, as shown in Figure~\ref{fig:corr_dep_backblaze}, time periods that are far from each other tend to have a larger difference in the distributions of the dependent variable than neighboring time periods do. 
For example, the distribution of the dependent variable in the 16th period has \texttt{medium} or \texttt{large} differences from the time periods that are far from the time period (i.e., periods 1-9 and 26-36) while only has \texttt{small} differences with its neighboring time periods (i.e., periods 10-15 and 17-25).
However, on the Google dataset, we observe that even the distribution between adjacent time periods can also have drastic changes.
As shown in Figure~\ref{fig:corr_dep_google}, two neighboring time periods (e.g., the periods 18 and 19) can have \texttt{medium} to \texttt{large} differences in their data distributions, while two time periods far away from each other (e.g., periods 1 and 26) can have \texttt{negligible} differences on the Google dataset.
We also observe that significant changes can happen between neighboring time periods on the Alibaba dataset, as shown in Figure~\ref{fig:corr_dep_alibaba}. 
Such results suggest that the obsolescence of AIOps models may not only be determined by time. 
Instead, we need more sophisticated methods to determine when to maintain AIOps models.

\textbf{Discussion on the seasonality in the datasets.}
Considering the two aforementioned values are the aggregation of various samples in a time period (i.e., the number of samples in a time period and the failure rate in a time period), the analysis of the trending may be impacted by the length of the time periods chosen for the aggregation.
In this work, we have chosen natural time periods (e.g., days, weeks, and months) that may represent the time patterns of practitioners' system operation activities.
Since our datasets are mostly monitoring data on server clusters, it is possible that the change of distribution is caused by repeating patterns in the natural periods (e.g., workloads in each week). 
Therefore, we also analyze the preliminary study results to evaluate the impact of temporal dependencies.
Similar to prior work~\cite{xue2018spatial}, we use autocorrelation to capture such temporal patterns within each dataset.
Autocorrelation is a mathematical representation that estimates the degree of similarity in a time series to a lagged version of itself.
The correlation coefficients of autocorrelation range in $[-1, 1]$. Higher positive values indicate stronger temporal dependencies, lower negative values indicate stronger diametrically opposed dependencies, and zero values suggest no temporal dependency. 
We apply autocorrelation to the failure rate and data size statistics on all three datasets. 
The statistics on the Google and Alibaba datasets show no significant correlations (i.e., correlation coefficients greater than the 95\% confidence interval) on any time lags. 
On the Backblaze dataset, we notice that the data size trend significantly correlates to the statistics from 1 to 3 months time lag. The failure rate trend shows moderate similarity to the data from 1 and 2 months time lag. However, no significant correlations are found in the data from longer time lags.
The seasonality on the Backblaze dataset confirms our choice of using one month as the length of time period. 
Overall, the findings suggest that no to little significant seasonality exists in the data size and failure rate statistics on the Google and Alibaba datasets, while on the Backblaze datasets, we found only short-term seasonality (i.e., 1 to 3 months compared with the total dataset with 36 months length).

\hfill \break
The results from our preliminary study above show that the studied operation data is constantly evolving. 
Therefore, it is necessary to perform model updates in order to ensure a satisfactory user experience. 
This point will be the focus of the remaining part of the paper.

\section{Experiment Design}
\label{sec:experimental_design}

In this section, we first describe our studied model update strategies and then describe how we build and evaluate AIOps models around these strategies.
\label{sec:model_approaches}

\subsection{Studied Model Update Strategies}
\begin{figure}[!t]
    \centering
    \includegraphics[width=0.68\textwidth]{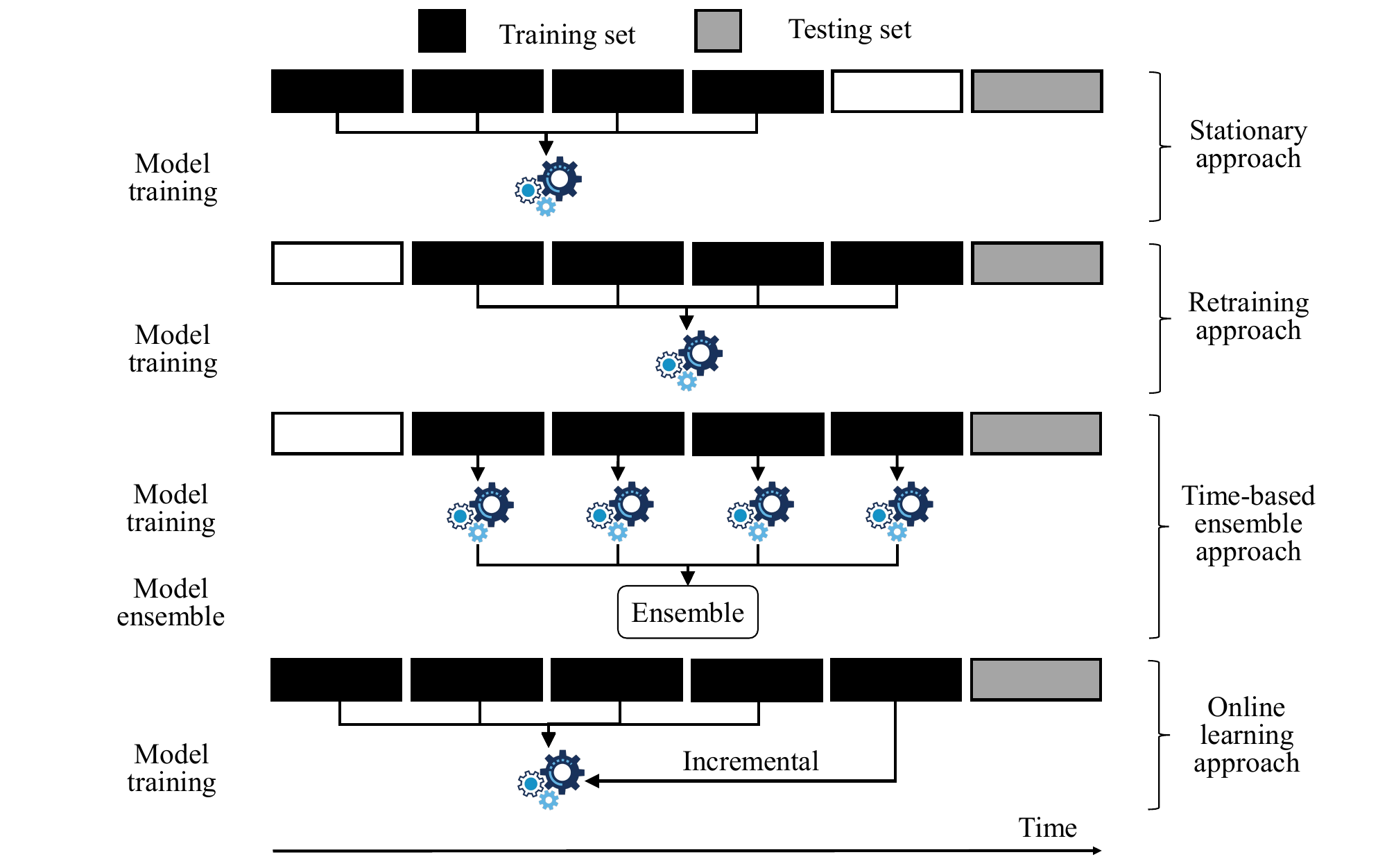}
    \caption{Illustration of different strategies for maintaining AIOps models. The illustration for the ``retraining approach'' represents both the periodical retraining and concept drift guided retraining strategies.}
    \label{fig:approach_overview}
\end{figure}

We consider five types of strategies for updating AIOps models: stationary (baseline), periodical retraining (baseline), concept drift guided model retraining, time-based ensemble, and online learning. Figure~\ref{fig:approach_overview} illustrates the five types of strategies.
For the concept drift guided retraining strategies, we consider three concept drift detection methods (DDM~\cite{gama2004learning}, PERM~\cite{harel2014concept}, and STEPD~\cite{nishida2007detect}).
For the time-based ensemble strategies, we consider AWE~\cite{wang2003mining} and SEA~\cite{street2001streaming} algorithms, as they are agnostic to the base classifiers. 
All strategies mentioned above can apply to any classification models used in AIOps solutions (except that the online learning strategies need the base models to be updatable); hence we build them above five types of ML models (i.e., LR, CART, RF, NN, and GBDT) used in prior works~\cite{mahdisoltani2017proactive, botezatu2016predicting, rosa2015predicting, elsayed2017learning} as their based models.
For the online learning strategies, we consider Hoeffding Tree (HT)~\cite{domingos2000mining}, Adaptive Random Forest (ARF)~\cite{gomes2017adaptive}, and Accuracy Updated Ensemble (AUE)~\cite{brzezinski2014reacting} algorithms.
We also consider two baseline models: 1) a stationary model baseline that builds the model on the initially available data and then never updates, and 2) a periodical retraining model that updates whenever training data from a new time period becomes available.

To evaluate different strategies for maintaining AIOps models, we divide the whole dataset into multiple time periods to simulate the scenario in which a chunk of data from a new time period becomes available each time.
We have 28 one-day time periods for the Google cluster trace data and 36 one-month time periods for the Backblaze disk stats data.

\subsubsection{Baseline Strategies}
Prior works usually use stationary models~\cite{elsayed2017learning, rosa2015predicting, botezatu2016predicting, mahdisoltani2017proactive, chen2019outage, xue2018spatial} or periodically retrain their models~\cite{lin2018predicting, li2020predicting} in the context of AIOps.
Accordingly, we consider two baseline strategies: stationary and periodical retraining.

\begin{itemize}
    \item \textbf{Stationary strategy (Stationary)}. 
    A stationary model is a model that is trained once and never updated.
    We train a model on the data from the first half of the time periods and apply it to all subsequent time periods. 
    It represents the scenario that a model is trained on the initially available data and never maintained when new data becomes available.
    
    \item \textbf{Periodical retraining strategy (Retrain)}.
    Periodically updating a model can maintain the model performance against the changes in the data distributions. 
    When the data from a new time period becomes available, we retrain a new model using the latest data (i.e., samples in the latest sliding window).
    Following prior work~\cite{lin2018predicting}, we use a sliding window instead of the entire historical data since the latter is usually unrealistic in the operational environment and may not increase the model performance as the data from old time periods may come from a different distribution~\cite{bifet2007learning, brzezinski2014reacting}.
    In our experiment, we fixed the sliding window size at half the total number of time periods (i.e., 14 days for the Google dataset, 18 months for the Backblaze dataset, and 4 weeks for the Alibaba dataset).
\end{itemize}

\subsubsection{Concept drift guided retraining strategies}
Prior works propose various methods~\cite{gama2004learning, harel2014concept, nishida2007detect} in detecting concept drift in mining stream data.
In this work, we apply these methods to detect changes in the operation data and update the models only when detecting drift in data distribution to preserve model performance while minimizing maintenance needs. 
Similar to the periodical retraining strategy, we set the length of the sliding window to half the number of time periods in the whole dataset (i.e., 14 days for the Google dataset, 18 months for the Backblaze dataset, and 4 weeks for the Alibaba dataset).
Algorithm~\ref{alg:concept_drift_experiment} describes how we retrain the models when an occurrence of concept drift is detected and how we evaluate the model performance.

\begin{algorithm}[!ht]
\SetAlgoLined
\KwIn{$N$: the total number of time periods\;
\hspace{3em}$[P_1,\dots,P_N]$: Data that is split into $N$ time periods
}
$M\leftarrow$ Model built on data $[P_1,\dots,P_{N/2}]$\;
\For{$i=N/2+1$ \KwTo $N-1$}{
    \If{concept drift detected on $P_i$}{
        $M\leftarrow$ New model built on data $[P_{i-N/2+1},\dots,P_{i}]$\;
    }
    Evaluate model performance on data $P_{i+1}$\;
}
\caption{The workflow of concept drift guided retraining strategies.}
\label{alg:concept_drift_experiment}
\end{algorithm}

\noindent\textbf{Concept drift detection methods}. 
We consider three methods for detecting concept drift and apply them in Algorithm~\ref{alg:concept_drift_experiment}.
The three methods are the Drift Detection Method (DDM)~\cite{gama2004learning}, the PERM concept drift detection method~\cite{harel2014concept}, and the STEPD concept drift detection method~\cite{nishida2007detect}.

\begin{itemize}
    \item \textbf{The Drift Detection Method (DDM)}. 
    Gama et al. propose the Drift Detection Method (DDM) based on the assumption that a significant increase in the testing error suggests a change in the data distribution (i.e., concept drift)~\cite{gama2004learning}.
    DDM maintains two registers $p_{min}$ and $s_{min}$, where $p$ is the error rates for different testing time periods and $s=sqrt(p(1-p)/n)$ is the standard deviation of $p$. 
    When the samples in a new time period become available, DDM calculates the new $p_i$ and $s_i$ using the error rate evaluated on the samples from the new time period, and updates $p_{min}$ and $s_{min}$ with $p_i$ and $s_i$ respectively when $p_i+s_i<p_{min}+s_{min}$. 
    Otherwise, if $p_i+s_i\geq p_{min}+3*s_{min}$, DDM judges that a concept drift is occurring.
    
    \item \textbf{The PERM concept drift detection method}. 
    Harel et al. propose the PERM method for detecting concept drift on streaming data based on the empirical loss of learning algorithms~\cite{harel2014concept}. 
    The intuition is that if no concept drift exists, the performance evaluation on an ordered split of data samples from two consecutive time periods should not deviate too much from the evaluation on shuffled splits. 
    Denote the original data from the two consecutive time periods as $S_{ord}$ and $S_{ord}'$, we generate $P$ random split $(S_i, S_i')$ of $S_{ord}\cup S_{ord}'$, then calculate the risks $\hat{R}_{S_i'}$ (i.e., the MSE loss function) of training a model on $S_i$ and testing it on $S_i'$.
    We then compare $\hat{R}_{S_i'}$ for $i$ in $P$ splits with the risk $\hat{R}_{ord}$ of training a model on $S_{ord}$ and testing it on $S_{ord}'$.
    Under given significant level $\delta$ and rate of change $\Delta$, if we confirm:
    \begin{equation}
        \frac{1+\sum_{i=1}^P1[\hat{R}_{ord}-\hat{R}_{S_i'}\leq \Delta]}{P+1}\leq \delta\textrm{,}
    \end{equation}
    then we suppose there is an occurrence of concept drift between the two time periods. Otherwise, there is no concept drift. 
    In our experiment, we choose parameters $\Delta=0$, $\delta=0.01$, and $P=100$, which are the same parameters as in the original paper~\cite{harel2014concept}.

    \item \textbf{The STEPD concept drift detection method}. 
    Nishida et al. propose a concept drift detection method that uses a statistical test of equal proportions (STEPD) to detect concept drift~\cite{nishida2007detect}.
    The STEPD algorithm assumes that the accuracy of a model for recent samples will be equal to the overall accuracy from the beginning of the learning if the target concept is stationary and a significant decrease in accuracy if the concept drifts.
    When data from a new time period becomes available, we calculate a two-proportion Z-test to compute the statistical difference between the model's prediction accuracy in the newest samples and old samples, as described below:
        \begin{equation}
            Z = \frac{(\hat{p_2}-\hat{p_1})-0}{\sqrt{\hat{p}(1-\hat{p})(\frac{1}{n_1}+\frac{1}{n_2})}}\textrm{,}
        \end{equation}
    where $\hat{p}$ is the overall prediction error rate, $\hat{p_1}$ and $n1$ are the prediction error rate and number of samples in the older time period, while $\hat{p_2}$ and $n2$ are the prediction error rate in the newest time period, respectively.
    When the $p$-value is less than $0.05$, we reject the null hypothesis (i.e., $\hat{p_1} - \hat{p_2}$ = 0) and consider the alternative that there is an occurrence of concept drift between the two time periods.
\end{itemize}

\subsubsection{Time-Based Ensemble Strategies}

The concept drift in operations data motivates us to consider training local base classifiers using the data from each relatively short time period and then combine these base classifiers as an ensemble~\cite{street2001streaming, wang2003mining, wang2010mining}.
Instead of retraining the whole model like in the above-mentioned periodical retraining and concept drift guided retraining strategies, time-based ensembles always train a smaller base classifier on the most recent samples and add it into the ensemble when certain criteria are met.
As a result, a time-based ensemble does not need to detect when the model needs to be updated and can exploit knowledge from even faraway historical time periods.
Specifically, we train separate base classifiers using samples from each time period, then combine these base classifiers to assemble an ensemble model for predicting future samples. 
Similar to the sliding window size for periodical retraining models, we fix the ensemble size as half the total amount of time periods. 
We consider the following two ensemble strategies in our experiment.

\begin{itemize}
    \item \textbf{Streaming Ensemble Algorithm (SEA)}. 
    Street and Kim propose an ensemble algorithm, namely SEA, for handling concept drift~\cite{street2001streaming} in the data mining context. 
    SEA combines multiple classifiers trained on different time periods using majority voting. 
    It keeps replacing the ``weakest'' base classifiers in the ensemble with new base classifiers according to a quality measure based on the base classifiers' predictions on the samples from the newly arrived time period. 
    The quality measure is a combined score that balances the accuracy and diversity of the base classifiers~\cite{street2001streaming}.
    Specifically, when samples from time period $i$ become available, these new samples are used to evaluate all base classifiers in the ensemble and a candidate classifier $C_{i-1}$ trained on the previous time period using the quality measure. If the quality of $C_{i-1}$ is better than a base classifier $E_j$ in the ensemble, then $E_j$ is replaced by $C_{i-1}$. Note that $C_{i-1}$ is automatically appended to the ensemble if the number of base classifiers in the ensemble has not reached the predefined size.
    Then, a new classifier $C_i$ is built on the samples from the time period $i$, and is saved as the candidate base classifier for the next time period.

    \item \textbf{Accuracy Weighted Ensemble (AWE)}.
    Another ensemble approach proposed by Wang et al.~\cite{wang2003mining, wang2010mining}, namely AWE, also aims at handling concept drift in streaming data.
    The algorithm of AWE is different from SEA in mainly two points:
    1) Instead of using a majority voting as in SEA, AWE applies a weighted ensembling approach that assigns weights (described below) on each base classifier in the ensemble. 
    2) When a new time period $i$ becomes available, SEA evaluates $C_{i-1}$ on the new time period to determine whether to add $C_{i-1}$ into the ensemble, while AWE evaluates $C_i$ on the new time period to determine whether to add $C_i$ into the ensemble.

    Each base classifier's weight in the ensemble is calculated based on its estimated prediction error on the testing data. Since the testing data is unseen at the training time, it assumes that the distribution of the most recent training data is close to the distribution of the testing data. Thus, the base classifiers' weights are approximated by computing their prediction error on the latest training data.
    Specifically, the weight of the $i$th base classifier is based on its mean square error on the latest training data as
    \begin{equation}
        \mathrm{MSE}_i = \frac{1}{N}\sum_{j=1}^N(y_j - \tilde{y}_j)^2\textrm{,}
    \end{equation}
    where $N$ is the number of samples in the latest training time period, $\tilde{y}_j$ is the predicted probability of the positive class, and $y_j$ is the observed class (1 for the positive class and 0 for the negative class). 
    The weight of the $i$th classifier is then defined as
    \begin{equation}
        w_i = \text{MSE}_r - \text{MSE}_i\textrm{,}
    \end{equation}
    where $\text{MSE}_r$ is the mean square error of a classifier that predicts randomly ($\text{MSE}_r = 0.25$). Base classifiers that produce an $\text{MSE}$ higher than $0.25$ are removed. 
    For the classifier trained on the latest training time period, we use 10-fold cross-validation to estimate its $\text{MSE}$.
\end{itemize}

\subsubsection{Online Learning Strategies}

The model update strategies mentioned above (i.e., stationary, periodical retraining, concept drift guided retraining, and time-based ensemble) are agnostic to the model type, which means they can use any supervised classification algorithms other than the five we evaluated.
However, these strategies require retraining each time the model is updated.
On the other hand, online learning strategies are more flexible in model updating but usually rely on specifically designed models for incremental model training.
We consider three online learning strategies as follows.

\begin{itemize}
    \item \textbf{Hoeffding Tree (HT)}. Proposed by Domingos et al.~\cite{domingos2000mining}, Hoefdding Tree (HT) is a widely adopted online prediction model that can incrementally learn from a massive data stream.
    HT uses Hoefdding bounds to ensure that the attributes chosen using a limited number of samples are the same as what would be chosen using infinite samples with a high probability.
    \item \textbf{Adaptive Random Forest (ARF)}. ARF~\cite{gomes2017adaptive} is a Random Forest algorithm designed for evolving data stream classification. It uses an algorithm to detect concept drift, 
    When the concept drift detection mechanism in ARF senses warning level signals, the model creates background trees that are not included in the prediction. These background trees are subsequently used to replace current trees in the forest if detecting concept drift. ARF also uses a weighted voting mechanism based on the testing accuracy to further cope with concept drift.
    \item \textbf{Accuracy Updated Ensemble (AUE)}. AUE is a time-based ensemble algorithm that uses online base classifiers to react to sudden drifts and can also gradually evolve with slowly changing concepts~\cite{brzezinski2014reacting}. The AUE algorithm is similar to the above-mentioned time-based ensemble models, except it uses online base classifiers (Hoeffding Tree) and updates the existing ensemble members with the newly arrived data samples.
\end{itemize}

\subsection{Building Predictive Models}
\label{sec:model_buidling}

\noindent\textbf{Models used in the stationary, periodical retraining, concept drift guided retraining, and time-based ensemble strategies}.
To ensure the generalizability of our experiment results, we choose a variety of models used in prior works~\cite{elsayed2017learning, mahdisoltani2017proactive, botezatu2016predicting} to predict disk failures on the Backblaze disk stats dataset and job failures on the Google and Alibaba cluster trace datasets. 
The list of models we select includes Logistic Regression (LR), Classification and Regression Trees (CART), Random Forest (RF), Neural Network (NN)~\footnote{Note that we use a vanilla NN with one hidden layer, same as prior works~\cite{rosa2015catching, mahdisoltani2017proactive}.}, and Gradient Boosting Decision Tree (GBDT).
We select these models as they are used in prior studies on the same datasets~\cite{elsayed2017learning, mahdisoltani2017proactive, botezatu2016predicting, rosa2015catching, rosa2015predicting}.
We choose the implementations in the \texttt{scikit-learn}\footnote{\url{https://scikit-learn.org/stable/}} Python package for all these models.
Due to the skewness of our datasets (i.e., only 1.5\% of the samples on the Google cluster trace dataset are related to job failures, and only 0.05\% of the samples in the Backblaze disk stats dataset are related to disk failures), we apply under-sampling on the training set for all of our models using a ratio of 1:10, same as in prior work~\cite{mahdisoltani2017proactive}.
For the Alibaba dataset, 34.5\% of the samples are related to job failures, and we have not applied downsampling to them.
The five types of models are applied to all the model update strategies except online learning.
We also apply a standard feature scaling transformation to avoid bias on features with large parameter value ranges.
As suggested in prior works~\cite{tantithamthavorn2018parameter, tantithamthavorn2016automated, song2013impact}, hyperparameter settings can significantly impact the performance of prediction models.
Therefore, we tune the hyperparameters of our studied models on the training data using a random search. 
We choose a random search instead of a grid search for our hyperparameter tuning as using random search is more efficient and can find models that are as good as or better than using a grid search~\cite{james2012random}.

\noindent\textbf{Models used in the online learning strategies.}
Online learning strategies typically have specially designed mechanisms to satisfy the need for constant model updating.
In this work, our studied HT model uses Hoefdding bounds to update the underlying decision tree model~\cite{domingos2000mining}.
The studied ARF model is a modified Random Forest algorithm that can learn incrementally~\cite{gomes2017adaptive}.
On the other hand, the studied AUE model is a time-based ensemble model that features online learning models as its base classifiers~\cite{brzezinski2014reacting}. We use the HT model as the base classifier for AUE, the same as in the original paper~\cite{brzezinski2014reacting}.
We choose the implementations in the \texttt{scikit-multiflow}\footnote{\url{https://scikit-multiflow.github.io/}} Python package for the HT and ARF models and implement the AUE model in Python~\footnote{AUE has an implementation in Java~\cite{brzezinski2014reacting} but no available Python implementation, so we implemented it in Python by ourselves for a fair comparison. The other two works do not mention the implementation of their algorithms, so we choose to use the popular \texttt{scikit-multiflow} Python package.}.

\subsection{Evaluating Model Performance}
\label{sec:model_evaluation}

In this work, we evaluate our studied model update strategies along three dimensions: performance, model updating cost, and stability.

\subsubsection{Performance.}
We evaluate the performance of our models using the AUC (Area Under the Receiver Operating Characteristic Curve) metric, which is a standard and widely used metric for evaluating machine learning models. 
AUC measures the model performance by calculating the area under the curve of true positive rate (TPR) against false positive rate (FPR) at different classification thresholds.
It evaluates a model's ability to discriminate between positive samples (e.g., failed jobs) and negative samples (e.g., normal jobs).
Prior work recommends the use of AUC over threshold-dependent metrics (e.g., precision and recall) when comparing model performance~\cite{tantithamthavorn2018experience}.
In our replication package, we also report other performance metrics (e.g., F1 score and MCC (Matthews Correlation Coefficient)). 
To mitigate the impact of random noise, we train and test each model with each strategy 100 times and calculate their average performance.

\subsubsection{Model updating cost}
While model update strategies can mitigate performance deterioration caused by the constant data evolution in operation datasets, significant computing efforts will be needed to update (e.g., by retraining) the model.
Once a model is updated, additional efforts and resources will also be needed to verify the quality of the model (i.e., testing), integrate the model into the production system (i.e., integration), deploy the system that includes the AIOps model to the production environment (i.e., deployment), and monitor the performance of the updated model in the production environment (i.e., monitoring). 
Therefore, AIOps solutions that require frequent updating may not be desirable~\cite{dang2019aiops,li2020predicting}. 
Accordingly, we define model updating cost from two perspectives: 1) the cost-effectiveness of model updates which quantifies the model performance gain normalized by the times of model update, and 2) the cost of computing resources in terms of the training and testing time.

\noindent\textbf{Cost-effectiveness analysis}.
In order to evaluate the benefits and costs of different strategies for determining when to update AIOps models, we perform cost-effectiveness analysis~\cite{weinstein1996cost} on each considered strategy.
In particular, for each strategy, we calculate the \textbf{effectiveness per unit of cost} (i.e., the \textbf{EC ratio})~\cite{EffectivenessPerCost} by Equation:
\begin{equation}
    EC = \frac{\textnormal{Performance improvement}}{\textnormal{Frequency of retrains}}\mathrm{,}
\end{equation}
where the performance improvement is represented by the percentage of AUC improvement over the stationary model, and the frequency of retrains is measured by the percentage of time periods needing updates. 
A higher EC value suggests the given strategy is more cost-effective.
We do not measure the EC ratio for the online learning strategies since the score is based on the performance improvement over the stationary strategies, while we do not have a corresponding base model in the stationary strategy for the online learning strategies.

\noindent\textbf{Training and inference cost}.
Considering the sheer volume of operation data and the need for real-time actions in specific scenarios, the computing cost should also be carefully inspected.
For example, training an LLM can cost millions of dollars, and the inference costs far exceed training costs when deploying a model at any reasonable scale, with ChatGPT costing over \$700,000 per day to operate in hardware inference costs~\cite{semianalysis}.
Therefore, we quantify the training and inference time of each model update strategy and model choice as one aspect of our evaluation and further estimate the compute cost in USD by referencing the cloud computing price per unit time~\cite{epochai}. 
We choose the \textrm{r6i.2xlarge} instance (3rd generation Intel XEON CPU with 64 GiB memory) on AWS with an on-demand hourly cost of \$0.504 to estimate the training and inference cost in dollars.
The CPU in this type of instance is a rough equivalent to our benchmark platform, and the memory is the minimum possible fit for running our experiments.
The training time measures the time needed for preprocessing, hyperparameter tuning, concept drift detection, and model fitting, and the inference time measures the total time needed to apply the models in predicting testing samples.
To mitigate measuring errors, we repeat each experiment 100 times on the same machines and choose the average. 
To further avoid bias in comparing the time efficiency of different strategies, we limit all models to only using a single CPU core.

\subsubsection{Stability.}
Due to the evolving nature of operation data, strategies that fail to maintain and update models timely may result in an inferior performance on future testing data and could further undermine the usefulness of the model.
Therefore, we include stability (i.e., to what extent the performance varies on testing data from different time periods) as one of the evaluation dimensions.
We measure the stability over each combination of update strategy and model choice by calculating the coefficient of variation (CV):
\begin{equation}
CV=\frac{\sigma}{\mu}\mathrm{,}
\end{equation}
where $\sigma$ is the standard deviation and $\mu$ is the mean value of the observations on the AUC performance from different testing time periods.
Similar to the experiment design above, we repeat the experiment 100 times on the same machines and calculate the average CV from the repetitions.

\subsubsection{Statistical ranking on the evaluation results.} 
For each of the above-mentioned evaluation metrics, we use the Scott-Knott test to rank the combinations of model update strategy and model choice statistically.
Scott-Knott is a clustering technique~\cite{scott1974cluster} that groups observations into statistically distinct groups using hierarchical clustering analysis. 
The observations within a group have no statistically significant difference (i.e., $p$-value $\geq$ 0.05), while the observations in different groups have a statistically significant difference (i.e., $p$-value $<$ 0.05).
In our case, the observations are the evaluated metric values (e.g., AUC) of the combinations of model update strategies and model choices in the repeated experiments.

\section{Experiment Results}
\label{sec:results}

We organize our experiment results by evaluating the studied model update strategies in three dimensions: performance, updating cost, and stability. 
We apply five types of models (i.e., LR, CART, RF, NN, and GBDT) to four strategies that incorporate regular ML models out of the five model update strategies we have: stationary, periodical retraining, concept drift guided retraining, and time-based ensemble. 
We also experiment with three types of models (i.e., HT, ARF, and AUE) for the online learning strategy.

\subsection{Performance}

\begin{figure}[!ht]
    \centering
    \subfloat[Google]{\label{fig:auc_trend_google}{
        \includegraphics[width=0.95\textwidth]{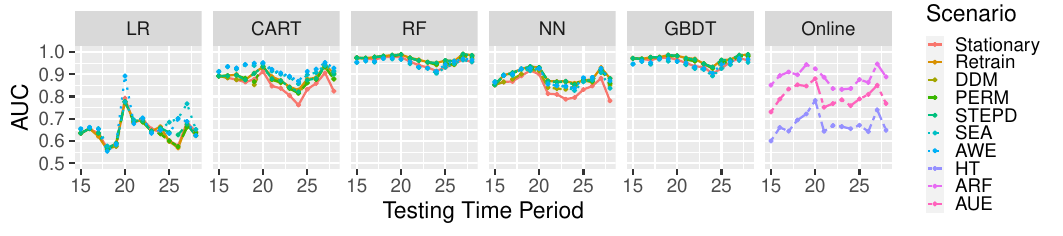}\hfill
    }}\hfill
    \subfloat[Backblaze]{\label{fig:auc_trend_backblaze}{
        \includegraphics[width=0.95\textwidth]{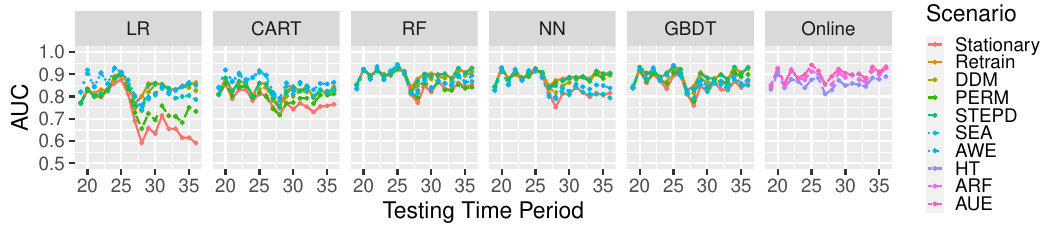}\hfill
    }}\hfill
    \subfloat[Alibaba]{\label{fig:auc_trend_alibaba}{
        \includegraphics[width=0.95\textwidth]{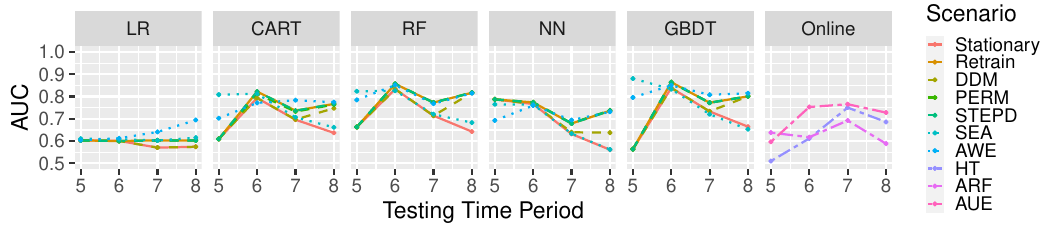}\hfill
    }}
    \caption{The AUC performance of different model update strategies in each testing period.}
    \label{fig:auc_trend}
\end{figure}

\begin{figure}[!ht]
    \centering
    \subfloat[Google]{\label{fig:auc_sk_google}{
        \includegraphics[width=0.95\textwidth]{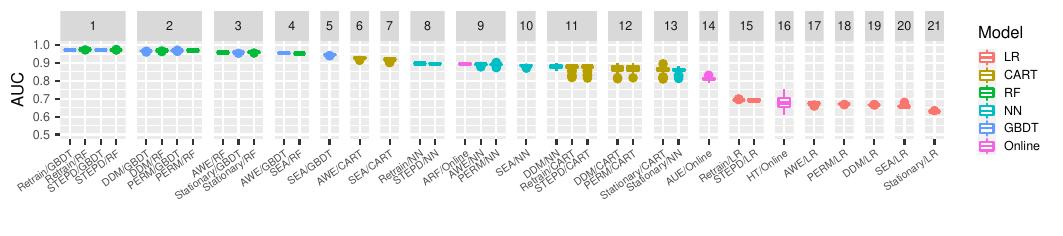}\hfill
    }}\hfill
    \subfloat[Backblaze]{\label{fig:auc_sk_backblaze}{
        \includegraphics[width=0.95\textwidth]{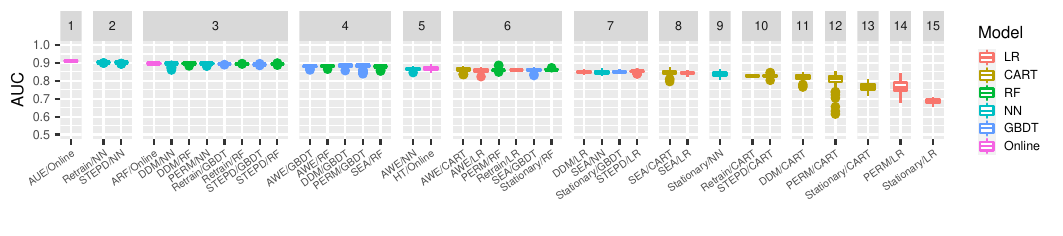}\hfill
    }}\hfill
    \subfloat[Alibaba]{\label{fig:auc_sk_alibaba}{
        \includegraphics[width=0.95\textwidth]{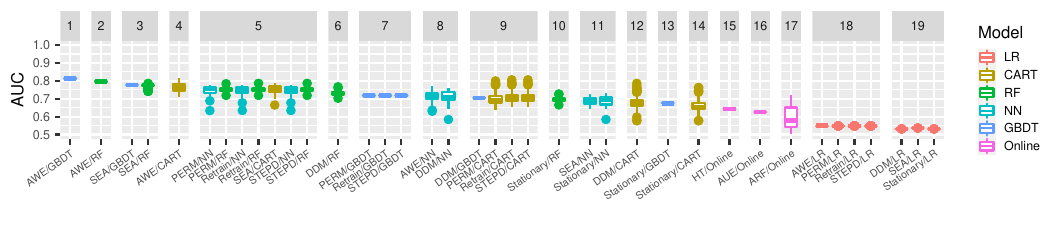}\hfill
    }}
    \caption{Scott-Knott test results of the AUC performance of different model update strategies.}
    \label{fig:auc_sk}
\end{figure}

Figure~\ref{fig:auc_trend} shows the AUC performance of each combination of update strategy and model choice on testing time periods.
Table~\ref{tab:performance_table} shows the overall AUC performance for the update strategies on each model choice and dataset. The aforementioned overall AUC performance is calculated on all the testing samples combined. 
Figure~\ref{fig:auc_sk} further shows our Scott-Knott ranking test results on the overall AUC performance of each combination of update strategy and model choice.

\begin{table}[!ht]
    \centering 
    \caption{The performance evaluation of different model maintenance strategies in terms of the overall AUC performance.}
    \label{tab:performance_table}
    \resizebox{\textwidth}{!}{
    \begin{threeparttable}
    \begin{tabular}{*{3}{c}*{10}{l}}
        \toprule
        \multirow{3}{*}{Item}& \multirow{3}{*}{Dataset}& \multirow{3}{*}{Model} & \multicolumn{10}{c}{Strategy}\\
        \cline{4-13}
        & & & \multirow{2}{*}{Stationary}& \multirow{2}{*}{Retrain}& \multicolumn{3}{c}{Detection}& \multicolumn{2}{c}{Emsemble}& \multicolumn{3}{c}{Online\tnote{2}}\\
        \cline{6-13}
        & & & & & DDM& PERM& STEPD& SEA& AWE& HT& ARF& AUE\\
        \midrule
        \multirow{15}{*}{AUC\tnote{1}}& \multirow{5}{*}{Google}& 
            LR & 0.63 (0.0\%) & 0.69 (10.3\%) & 0.66 (5.6\%) & 0.67 (6.7\%) & 0.69 (10.0\%) & 0.66 (4.5\%) & 0.67 (7.1\%) &\multirow{5}{*}{0.68}& \multirow{5}{*}{0.89}& \multirow{5}{*}{0.81}\\ 
        & & CART & 0.86 (0.0\%) & 0.88 (2.0\%) & 0.87 (1.0\%) & 0.87 (1.0\%) & 0.88 (1.9\%) & 0.92 (6.9\%) & 0.93 (7.8\%) \\ 
        & & RF & 0.96 (0.0\%) & 0.97 (1.5\%) & 0.97 (1.0\%) & 0.97 (1.1\%) & 0.97 (1.5\%) & 0.95 (-0.8\%) & 0.96 (-0.2\%) \\ 
        & & NN & 0.86 (0.0\%) & 0.90 (4.6\%) & 0.88 (2.4\%) & 0.89 (4.0\%) & 0.90 (4.4\%) & 0.88 (3.0\%) & 0.89 (3.9\%) \\ 
        & & GBDT & 0.96 (0.0\%) & 0.97 (1.5\%) & 0.97 (1.0\%) & 0.97 (1.1\%) & 0.97 (1.5\%) & 0.95 (-1.3\%) & 0.95 (-0.4\%) \\
        \cline{2-13}
        & \multirow{5}{*}{Backblaze}& 
            LR & 0.69 (0.0\%) & 0.86 (25.4\%) & 0.85 (23.9\%) & 0.76 (10.9\%) & 0.85 (24.5\%) & 0.84 (23.0\%) & 0.86 (25.1\%) &\multirow{5}{*}{0.87}& \multirow{5}{*}{0.90}& \multirow{5}{*}{0.91}\\ 
        & & CART & 0.77 (0.0\%) & 0.83 (7.9\%) & 0.82 (6.9\%) & 0.80 (4.9\%) & 0.83 (8.0\%) & 0.84 (10.1\%) & 0.86 (12.5\%) \\ 
        & & RF & 0.86 (0.0\%) & 0.90 (3.8\%) & 0.90 (4.0\%) & 0.86 (-0.3\%) & 0.90 (3.8\%) & 0.88 (2.0\%) & 0.88 (2.3\%) \\ 
        & & NN & 0.84 (0.0\%) & 0.90 (7.8\%) & 0.90 (6.9\%) & 0.90 (6.9\%) & 0.90 (7.6\%) & 0.85 (1.3\%) & 0.87 (3.4\%) \\ 
        & & GBDT & 0.85 (0.0\%) & 0.89 (5.0\%) & 0.88 (4.1\%) & 0.88 (3.8\%) & 0.89 (4.8\%) & 0.86 (1.2\%) & 0.88 (3.6\%) \\ 
        \cline{2-13}
        & \multirow{5}{*}{Alibaba}& 
            LR & 0.53 (0.0\%) & 0.55 (3.1\%) & 0.53 (0.0\%) & 0.55 (3.1\%) & 0.55 (3.1\%) & 0.54 (0.9\%) & 0.55 (3.4\%) &\multirow{5}{*}{0.64}& \multirow{5}{*}{0.60}& \multirow{5}{*}{0.62}\\ 
        & & CART & 0.66 (0.0\%) & 0.71 (6.2\%) & 0.68 (2.4\%) & 0.70 (5.4\%) & 0.71 (6.2\%) & 0.75 (13.0\%) & 0.76 (14.7\%) \\ 
        & & RF & 0.70 (0.0\%) & 0.75 (7.8\%) & 0.73 (5.0\%) & 0.75 (7.9\%) & 0.75 (7.8\%) & 0.77 (11.1\%) & 0.80 (14.3\%) \\ 
        & & NN & 0.69 (0.0\%) & 0.75 (8.5\%) & 0.71 (3.2\%) & 0.75 (8.7\%) & 0.75 (8.5\%) & 0.69 (-0.1\%) & 0.71 (3.5\%) \\ 
        & & GBDT & 0.67 (0.0\%) & 0.72 (6.5\%) & 0.70 (4.4\%) & 0.72 (6.5\%) & 0.72 (6.5\%) & 0.78 (15.4\%) & 0.81 (20.6\%) \\ 
        \bottomrule
    \end{tabular}
    \begin{tablenotes}
    \item[1] The number in brackets shows the AUC performance improvement compared with corresponding stationary models.
    \item[2] We skip the calculation of performance improvement for online models as they do not have corresponding stationary models.
    \end{tablenotes}
    \end{threeparttable}
    }
\end{table}

\textbf{Periodical retraining, concept drift guided retraining, and time-based ensemble strategies all achieve better performance than the stationary strategy, while different model update strategies lead to diverse model performance}.
We observe that different strategies for training AIOps models can lead to significant performance differences.
For example, the CART model trained on the Backblaze dataset achieves an overall AUC of $0.86$ using the AWE ensemble strategy and $0.83$ using periodical model retraining or the STEPD concept drift detection strategy. In contrast, the stationary model only achieves an overall AUC of $0.77$.
As shown in Figure~\ref{fig:auc_sk}, we also rank the AUC performance with our Scott-Knott test. 
It is worth noting that the stationary strategies always sit in the worst-performed group, except for the GBDT and RF models on the Google dataset, where time-based ensemble strategies (SEA and AWE) perform equally or worse than the stationary strategy.
Overall, the periodical retraining strategy provides 1.5\% to 10.3\% performance improvement on the Google dataset, 3.8\% to 25.4\% improvement on the Backblaze dataset, and 3.1\% to 8.5\% improvement on the Alibaba dataset over the stationary models. 
Although having less frequent retraining frequency, the concept drift guided retraining strategies can also provide comparable performance improvement to the periodical retraining strategies. We observe a 1.0\% to 10.0\% performance improvement for the Google dataset, 3.8\% to 24.5\% improvement for the Backblaze dataset, and 2.4\% to 8.7\% improvement for the Alibaba dataset over the stationary models.
For the time-based ensemble strategies, we also observe significant performance improvement, achieving a 3.0\% to 7.8\% increase on the Google dataset, 1.2\% to 25.1\% increase on the Backblaze dataset, and 0.9\% to 20.6\% increase on the Alibaba dataset over the stationary strategy.

\textbf{In general, periodical retraining strategies achieve the best performance among all the evaluated model update strategies, while concept drift guided retraining strategies can usually achieve as good performance}.
Our Scott-Knott analysis of the AUC performance (Figure~\ref{fig:auc_sk}) shows that the periodical retraining strategy achieves the best performance for $4$ out of the $5$ models on both the Google and Backblaze datasets and $3$ out of the $5$ models on the Alibaba dataset, while the ensemble strategies achieve better performance on the remained scenarios.
The Scott-Knott analysis results also show that concept drift guided retraining strategies achieve a performance level that has no statistical difference from the performance of the periodical retraining strategy for most of the scenarios.
For example, on the Google dataset, the STEPD concept drift guided retraining strategy ranks in the same group as the periodical retraining strategy for all $5$ models (i.e., LR, CART, RF, NN, and GBDT).
On the Backblaze dataset, the STEPD strategy also ranks in the same group as the periodical retraining strategy for $4$ of the $5$ models (i.e., CART, RF, NN, and GBDT).
Similarly, on the Alibaba dataset, the STEPD and PERM concept drift guided retraining strategies also rank in the same group as the periodical retraining strategy for all $5$ models.

\textbf{Time-based ensemble strategies achieve promising results for certain scenarios (tree-based models especially) while showing inferior performance than model retraining strategies in other scenarios}.
Prior works~\cite{wang2003mining, wang2010mining, street2001streaming} show that time-based ensemble strategies outperform training a single model.
For example, researchers found the AWE ensemble built on several models outperforms the corresponding single classifier on both synthetic and real-life stream data~\cite{wang2003mining}.
As shown in Figure~\ref{fig:auc_sk}, the time-based ensemble strategies achieve even higher performance than the periodical retraining strategy in several scenarios in our case study.
We observe both time-based ensemble strategies rank in the first or the second group for the CART model on all three datasets.
In addition, both time-based ensemble strategies also rank in the first or second group for the GBDT and RF models on the Alibaba dataset.
In particular, for the CART model on the Google dataset, the SEA and AWE time-based ensemble strategies achieve a 4.5\% and a 5.7\% overall performance increase compared to the best model retraining strategy (including concept drift guided retraining and periodical retraining strategy) in terms of the overall AUC, as shown in Table~\ref{tab:performance_table}.
However, in the other specific scenarios, the time-based ensemble strategies achieve poorer performance, especially on the NN model, where both ensemble models achieve similar performance to the stationary model.
In other scenarios, the time-based ensemble strategies achieve comparable performance to concept drift guided retraining strategies.
Part of the inferior results may be explained by the fact that these models already have internal ensemble mechanisms (i.e., RF and GBDT use bagging and boosting, respectively, and NN uses a strategy similar to the stacking ensemble strategy), thus adding an extra layer of ensemble process to these models may not always improve their performance. 

\textbf{Online learning models' performance is mixed and significantly depends on characteristics of the datasets}.
The performance of online learning strategies varies from dataset to dataset.
The online learning models rank among the worst-performing strategies on the Google and Alibaba datasets while ranking among the top-performing strategies on the Backblaze dataset. 
For example, the AUE and HT online learning strategies rank in the 14th and 16th performance groups in terms of AUC on the Google dataset, respectively (shown in Figure~\ref{fig:auc_sk_google}).
For the Alibaba dataset, online learning strategies also show inferior performance, with all three strategies ranking in the 15th to the 17th performance groups, as shown in Figure~\ref{fig:auc_sk_alibaba}.
In contrast, the online learning models are among the best-performing strategies on the Backblaze dataset, with the AUE online learning model being the best-performing strategy and the ARF online learning model ranking in the 3rd group for the Backblaze dataset, as shown in Figure~\ref{fig:auc_sk_backblaze}.
We speculate that the different characteristics of the datasets cause such significantly different results. 
For instance, as shown in Figure~\ref{fig:failure_rates}, the distribution of the target variable (i.e., the failure rate) in the Google and Alibaba datasets fluctuates rapidly from one time period to another, which may impair the performance of the online models that build up incrementally. In comparison, the distribution of the target variable in the Backblaze dataset is evolving gradually and is more predictable. 
Future work is needed to investigate the appropriate application scenarios for online learning models.

\subsection{Model updating cost}
\begin{figure}[!ht]
    \centering
    \subfloat[Google]{\label{fig:ec_sk_google}{
        \includegraphics[width=0.95\textwidth]{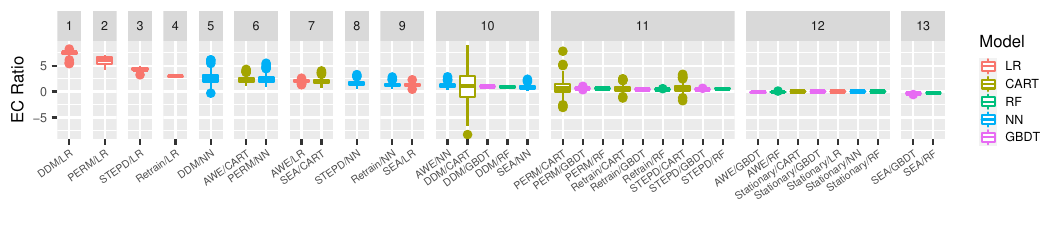}\hfill
    }}\hfill
    \subfloat[Backblaze]{\label{fig:ec_sk_backblaze}{
        \includegraphics[width=0.95\textwidth]{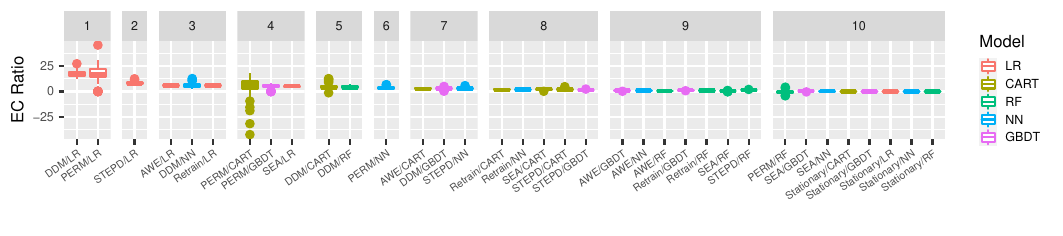}
    }}\hfill
    \subfloat[Alibaba]{\label{fig:ec_sk_alibaba}{
        \includegraphics[width=0.95\textwidth]{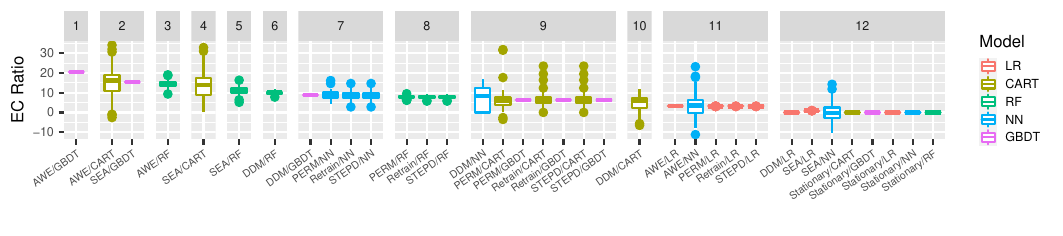}\hfill
    }}
    \caption{Scott-Knott test results of the EC ratio of different model update strategies.}
    \label{fig:ec_sk}
\end{figure}

\begin{figure}[!ht]
    \centering
    \subfloat[Google]{\label{fig:time_sk_google}{
        \includegraphics[width=0.95\textwidth]{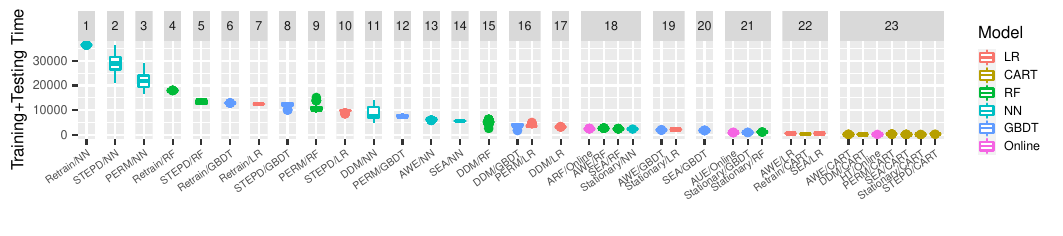}\hfill
    }}\hfill
    \subfloat[Backblaze]{\label{fig:time_sk_backblaze}{
        \includegraphics[width=0.95\textwidth]{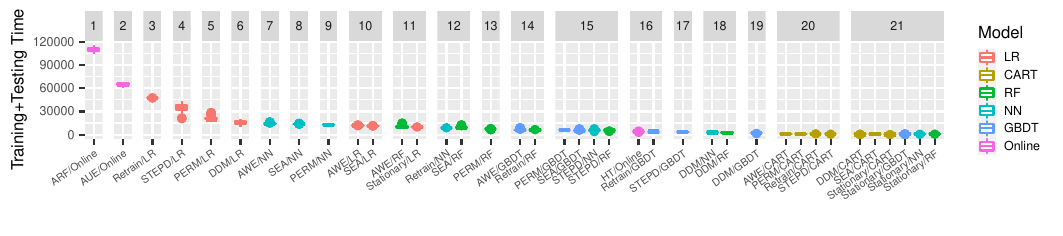}\hfill
    }}\hfill
    \subfloat[Alibaba]{\label{fig:time_sk_alibaba}{
        \includegraphics[width=0.95\textwidth]{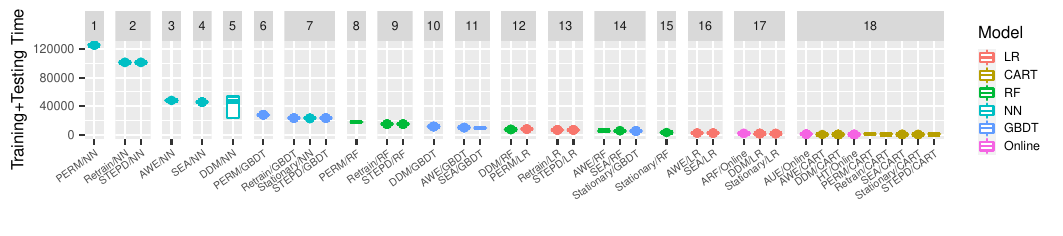}\hfill
    }}
    \caption{Scott-Knott test results of the training and testing time of different model update strategies.}
    \label{fig:time_sk}
\end{figure}

Table~\ref{tab:ec_table} shows the EC ratio of the combinations of our studied model update strategies and models. 
Figure~\ref{fig:ec_sk} shows the Scott-Knott test results of ranking the combinations into statistically distinct groups based on the EC ratios in 100 rounds of the experiment. 
Table~\ref{tab:effort_table} shows the training and testing time of each combination of model update strategy and model choice on all the training and testing time periods combined.
Figure~\ref{fig:time_sk} shows the Scott-Knott test results of ranking the combinations of model update strategy and model choice based on their training and testing time combined in 100 rounds of experiment.
Table~\ref{tab:cost_estimation_table} further shows the average cost estimation in dollar value for training and testing different strategies.

\begin{table}[!ht]
    \caption{The model updating cost evaluation of different model maintenance strategies in terms of the EC ratio.}
    \label{tab:ec_table}
    \begin{threeparttable}
    \begin{tabular}{*{3}{c}*{7}{r}}
        \toprule
        \multirow{3}{*}{Item}& \multirow{3}{*}{Dataset}& \multirow{3}{*}{Model} & \multicolumn{7}{c}{Strategy}\\
        \cline{4-10}
        & & & \multirow{2}{*}{Stationary}& \multirow{2}{*}{Retrain}& \multicolumn{3}{c}{Detection}& \multicolumn{2}{c}{Emsemble}\\
        \cline{6-10}
        & & & & & DDM& PERM& STEPD& SEA& AWE\\
        \midrule
        \multirow{15}{*}{EC ratio~\tnote{1}}& \multirow{5}{*}{Google}&
            LR & - (0) & 2.96 (13) & 7.43 (2) & 5.99 (4) & 4.28 (8) & 1.29 (13) & 2.04 (13) \\ 
        & & CART & - (0) & 0.57 (13) & 0.95 (3) & 0.66 (5) & 0.72 (10) & 1.97 (13) & 2.24 (13) \\ 
        & & RF & - (0) & 0.43 (13) & 0.94 (3) & 0.62 (7) & 0.55 (10) & -0.23 (13) & -0.07 (13) \\ 
        & & NN & - (0) & 1.32 (13) & 2.62 (2) & 2.33 (6) & 1.60 (10) & 0.87 (13) & 1.13 (13) \\ 
        & & GBDT & - (0) & 0.43 (13) & 0.98 (3) & 0.61 (6) & 0.46 (12) & -0.37 (13) & -0.12 (13) \\ 
        \cline{2-10}
        & \multirow{5}{*}{Backblaze}&
            LR & - (0) & 5.65 (17) & 17.34 (5) & 16.83 (1) & 8.14 (11) & 5.11 (17) & 5.58 (17) \\ 
        & & CART & - (0) & 1.76 (17) & 4.41 (6) & 4.85 (3) & 2.04 (15) & 2.26 (17) & 2.80 (17) \\ 
        & & RF & - (0) & 0.85 (17) & 4.07 (3) & -0.59 (1) & 1.24 (11) & 0.45 (17) & 0.51 (17) \\ 
        & & NN & - (0) & 1.74 (17) & 5.81 (4) & 3.43 (7) & 2.80 (10) & 0.29 (17) & 0.75 (17) \\ 
        & & GBDT & - (0) & 1.11 (17) & 2.81 (5) & 4.94 (2) & 1.53 (12) & 0.27 (17) & 0.81 (17) \\ 
        \cline{2-10}
        & \multirow{5}{*}{Alibaba}&
            LR & - (0) & 3.12 (3) & 0.00 (0) & 3.12 (3) & 3.12 (3) & 0.91 (3) & 3.38 (3) \\ 
        & & CART & - (0) & 6.31 (3) & 4.90 (1) & 6.15 (3) & 6.31 (3) & 13.34 (3) & 15.01 (3) \\ 
        & & RF & - (0) & 7.78 (3) & 10.06 (1) & 7.87 (3) & 7.78 (3) & 11.12 (3) & 14.34 (3) \\ 
        & & NN & - (0) & 8.56 (3) & 6.47 (1) & 8.76 (3) & 8.56 (3) & -0.00 (3) & 3.65 (3) \\ 
        & & GBDT & - (0) & 6.47 (3) & 8.85 (1) & 6.47 (3) & 6.47 (3) & 15.36 (3) & 20.59 (3) \\ 
        \bottomrule
    \end{tabular}
    \begin{tablenotes}
    \item[1] The number in brackets shows the counts of model retraining for each of the model maintenance strategies.
    \end{tablenotes}
    \end{threeparttable}
\end{table}

\textbf{The concept drift guided retraining strategies are among the most cost-effective and require fewer model updates}.
We observe that using concept drift guided retraining strategies can significantly reduce the model retraining frequency compared with the periodical retraining strategy (Table~\ref{tab:ec_table}).
For example, on the Backblaze dataset, the DDM, PERM, and STEPD concept drift guided retraining strategies only require 4, 7, and 10 times of retraining for the NN model, respectively. 
In contrast, the periodical retraining strategy retrains the same model type 17 times.
Overall, the concept drift guided retraining models achieve the highest EC ratio on the three datasets, except that the two time-based ensemble strategies (SEA and AWE) prevail in EC ratios for the CART model on the three datasets and for the RF and GBDT models on the Alibaba dataset.
Specifically, the DDM concept drift guided retraining strategy achieves the highest EC ratio on the Google dataset for 4 out of the $5$ models (i.e., LR, RF, NN, and GBDT); while on the Backblaze dataset, the PERM concept drift guided retraining strategy has the highest EC ratio for $3$ models (i.e., LR, CART, and GBDT) and the DDM strategy has the highest EC ratio for the other $2$ models (i.e., RF and NN).
The Alibaba dataset only has 3 model update periods and can only reflect short-term performance for the concept drift guided retraining strategies. 
We observe that the PERM and STEPD strategies retrain on all 3 testing periods and obtain similar EC ratios to the periodical retraining strategies. 
However, the DDM retraining strategy detects effective retraining points, only retraining once but obtaining similar EC performance to the periodical retraining strategy, except for the LR model, in which the DDM strategy detects no drift.
In addition, we observe the ensemble models tend to have a better EC ratio on the Alibaba dataset than on the other two datasets due to the large amount of performance improvement.
The Scott-Knott test results of the EC ratios (Figure~\ref{fig:ec_sk}) also show that the DDM strategy ranks in the first group for 4 of 5 models on the Google dataset and 3 of 5 models on the Backblaze dataset.
On the Alibaba dataset, the ensemble model has the best EC ratio, but the DDM strategy still manages to rank higher than the periodical retraining strategy on the GBDT and RF models.
Model update strategies other than the concept drift guided retraining strategies (i.e., periodical retraining and time-based ensemble) need to update the model each time the data in a new time period becomes available, which may not be favorable in field deployment.
Our results indicate that selectively maintaining the models based on concept drift detection could reduce the cost of model updating while cost-effectively preserving model performance.

\begin{table}[!ht]
    \centering 
    \caption{The evaluation of different strategies in updating AIOps models in terms of efforts. The total training time and testing time are calculated as the sum of the training/testing time for each round, including the time taken to detect concept drift.}
    \label{tab:effort_table}
    \resizebox{\textwidth}{!}{
    \begin{tabular}{*{3}{c}*{10}{r}}
        \toprule
        \multirow{3}{*}{Item}& \multirow{3}{*}{Dataset}& \multirow{3}{*}{Model} & \multicolumn{10}{c}{Strategy}\\
        \cline{4-13}
        & & & \multirow{2}{*}{Stationary}& \multirow{2}{*}{Retrain}& \multicolumn{3}{c}{Detection}& \multicolumn{2}{c}{Emsemble}& \multicolumn{3}{c}{Online}\\
        \cline{6-13}
        & & & & & DDM& PERM& STEPD& SEA& AWE& HT& ARF& AUE\\
        \midrule
        \multirow{15}{*}{\specialcell{Training\\time\\(seconds)}}& \multirow{5}{*}{Google}& 
            LR & 2,037 & 12,517 & 3,052 & 3,602 & 9,553 & 415 & 453 &\multirow{5}{*}{20}& \multirow{5}{*}{1,344}& \multirow{5}{*}{332}\\ 
        & & CART & 19 & 285 & 79 & 183 & 226 & 51 & 68 \\ 
        & & RF & 1,047 & 17,974 & 5,083 & 10,649 & 13,400 & 2,252 & 2,412 \\
        & & NN & 2,262 & 36,419 & 8,937 & 21,894 & 28,985 & 5,415 & 6,069 \\
        & & GBDT & 874 & 12,859 & 3,729 & 7,456 & 12,071 & 1,657 & 1,870 \\
        \cline{2-13}
        & \multirow{5}{*}{Backblaze}& 
            LR & 9,874 & 47,487 & 15,714 & 21,268 & 33,887 & 11,192 & 11,667 &\multirow{5}{*}{67}& \multirow{5}{*}{4,858}& \multirow{5}{*}{1,467}\\ 
        & & CART & 53 & 865 & 362 & 974 & 776 & 158 & 498 \\
        & & RF & 307 & 6,125 & 2,084 & 7,010 & 5,178 & 5,029 & 5,880 \\
        & & NN & 437 & 8,720 & 2,604 & 12,403 & 5,738 & 12,744 & 13,611 \\ 
        & & GBDT & 280 & 3,728 & 1,685 & 5,584 & 2,971 & 3,795 & 4,600 \\
        \cline{2-13}
        & \multirow{5}{*}{Alibaba}& 
            LR & 1,417 & 6,581 & 1,417 & 7,798 & 6,581 & 2,115 & 2,174 &\multirow{5}{*}{87}& \multirow{5}{*}{1,548}& \multirow{5}{*}{486}\\ 
        & & CART & 87 & 423 & 193 & 482 & 423 & 154 & 160 \\ 
        & & RF & 2,995 & 14,783 & 7,239 & 17,486 & 14,785 & 5,468 & 5,607 \\ 
        & & NN & 22,984 & 101,350 & 38,970 & 125,480 & 101,351 & 45,693 & 47,925 \\ 
        & & GBDT & 5,098 & 23,221 & 11,388 & 27,923 & 23,223 & 9,511 & 9,857 \\ 
        \midrule
        \multirow{15}{*}{\specialcell{Testing\\time\\(seconds)}}& \multirow{5}{*}{Google}& 
            LR & 0.3 & 0.2 & 0.2 & 0.3 & 0.2 & 10.7 & 10.7 &\multirow{5}{*}{43.2}& \multirow{5}{*}{958.5}& \multirow{5}{*}{456.7}\\
        & & CART & 0.4 & 0.3 & 0.3 & 0.3 & 0.3 & 10.4 & 10.1 \\
        & & RF & 3.6 & 4.7 & 4.7 & 4.6 & 4.7 & 85.5 & 85.9 \\
        & & NN & 0.6 & 0.8 & 0.7 & 0.8 & 0.8 & 24.7 & 24.9 \\
        & & GBDT & 1.8 & 2.3 & 2.5 & 2.2 & 2.3 & 39.8 & 41.9 \\
        \cline{2-13}
        & \multirow{5}{*}{Backblaze}& 
            LR & 11.9 & 11.8 & 11.4 & 11.4 & 11.2 & 189.4 & 192.3 &\multirow{5}{*}{3,760.9}& \multirow{5}{*}{10,5523.3}& \multirow{5}{*}{63,373.3}\\
        & & CART & 16.3 & 16.4 & 15.9 & 15.7 & 15.5 & 285.6 & 289.0 \\ 
        & & RF & 251.5 & 238.4 & 216.5 & 357.9 & 248.5 & 3989.1 & 4313.4 \\ 
        & & NN & 59.8 & 33.6 & 38.0 & 36.0 & 33.3 & 771.6 & 764.0 \\ 
        & & GBDT & 174.3 & 87.1 & 117.0 & 114.6 & 87.5 & 1557.1 & 1598.8 \\ 
        \cline{2-13}
        & \multirow{5}{*}{Alibaba}& 
            LR & 0.2 & 0.2 & 0.2 & 0.2 & 0.2 & 0.7 & 0.7 &\multirow{5}{*}{42.2}& \multirow{5}{*}{200.9}& \multirow{5}{*}{164.2}\\
        & & CART & 0.3 & 0.2 & 0.2 & 0.2 & 0.2 & 0.9 & 0.9 \\ 
        & & RF & 2.6 & 3.0 & 2.9 & 3.0 & 2.9 & 11.9 & 12.9 \\ 
        & & NN & 0.9 & 0.7 & 0.7 & 0.7 & 0.7 & 3.5 & 3.1 \\ 
        & & GBDT & 3.8 & 2.3 & 3.1 & 2.3 & 2.3 & 7.5 & 6.8 \\ 
        \bottomrule
    \end{tabular}
    }
\end{table}

\begin{table}[!ht]
    \centering 
    \caption{The cost estimation in dollars for running different model maintenance strategies in a single run. For strategies other than online learning methods, we take the average value from running five different models.}
    \label{tab:cost_estimation_table}
    \resizebox{\textwidth}{!}{
        \begin{tabular}{*{2}{c}*{10}{r}}
            \toprule
            \multirow{3}{*}{Item}& \multirow{3}{*}{Dataset}& \multicolumn{10}{c}{Strategy}\\
            \cline{3-12}
            & & \multirow{2}{*}{Stationary}& \multirow{2}{*}{Retrain}& \multicolumn{3}{c}{Detection}& \multicolumn{2}{c}{Emsemble}& \multicolumn{3}{c}{Online}\\
            \cline{5-12}
            & & & & DDM& PERM& STEPD& SEA& AWE& HT& ARF& AUE\\
            \midrule
            \multirow{3}{*}{\specialcell{Training\\cost (\$)}}& Google& 
            629 & 8,069 & 2,105 & 4,413 & 6,475 & 987 & 1,096 & 10& 675& 167\\ 
            \cline{3-12}
            & Backblaze& 
            1,104 & 6,746 & 2,263 & 4,762 & 4,894 & 3,191 & 3,506 & 34 & 2,439& 736\\ 
            \cline{3-12}
            & Alibaba& 
            3,284 & 14,753 & 5,968 & 18,060 & 14,754 & 6,345 & 6,625 & 44 & 777 & 244\\ 
            \midrule
            \multirow{3}{*}{\specialcell{Testing\\cost (\$)}}& Google& 
            1 & 1 & 1 & 1 & 1 & 17 & 17 & 22 & 481 & 229\\
            \cline{3-12}
            & Backblaze& 
            52 & 39 & 40 & 54 & 40 & 584 & 613 & 1,888 & 52,973 & 31,813\\
            \cline{3-12}
            & Alibaba& 
            1 & 1 & 1 & 1 & 1 & 2 & 2 & 21 & 101& 82\\
            \bottomrule
        \end{tabular}
    }
\end{table}

\textbf{Overall, the online learning and periodical retraining strategies usually have the highest time consumption for model training and testing, while the concept drift guided retraining and time-based ensemble strategies typically have the lowest time consumption}.
As shown in our Scott-Knott test results (Figure~\ref{fig:time_sk}) that rank the different combinations of model update strategies and model choices, the periodical retraining strategies of the NN, RF, and GBDT models take the longest time to finish a round of training and testing on the Google dataset.
On the Backblaze dataset, two online learning strategies (i.e., ARF and AUE) and the periodical retraining strategies for the LR model take the longest to train and test.
Due to the smaller amount of time periods on the Alibaba dataset, the model type becomes the dominant factor for training and testing time, and the strategies for the NN model take the longest to run.
However, in most cases, the concept drift guided retraining and time-based ensemble strategies take the shortest time to train and test.
On the Google datasets, the AWE and SEA time-based ensemble strategies are the fastest strategies other than the stationary models for all 5 types of models (i.e., LR, CART, RF, NN, and GBDT).
On the Backblaze dataset, the DDM concept drift guided retraining strategy ranks in the fastest groups, excluding the stationary models for 4 out of 5 models (i.e., CART, RF, NN, and GBDT).
On the Alibaba dataset, the two time-based ensemble strategies rank in the fastest groups other than the stationary models for 3 types of models (i.e., RF, CART, and GBDT), while the DDM concept-drift detection strategy ranks in the first place for the 2 remaining models (i.e., LR and NN).
Table~\ref{tab:cost_estimation_table} further shows the average cost estimation in dollars on five different model types for each model update strategy (except for the online learning strategy, in which we use the nominal value as it does not allow different model types).
We observe that the DDM concept drift detection strategy and the two time-based ensemble strategies have the lowest model training cost other than the stationary and online learning strategies, which matches our observations on time consumption.
Online learning can also have a short training time, but the performance varies significantly, with the HT online learning strategy being worse than the stationary CART model on both the Google and Alibaba datasets while having a longer training time.
We also observe that although time-based ensemble strategies have a higher cost for inference, the volume is still a fraction compared to the training cost.
Nevertheless, the time consumption and related computing cost are highly related to the choice of model and dataset. Hence, we advise practitioners to experiment with multiple models.

\textbf{The concept drift guided retraining, time-based ensemble, and online learning strategies can all reduce the total training time needed to update the models than the periodical retraining strategy}.
As shown in Table~\ref{tab:effort_table}, we observe that the DDM and STEPD concept drift guided retraining strategies show a significant reduction in the training time (i.e., 54.8\% to 75.6\% and 10.2\% to 34.2\%, respectively) compared with the periodical retraining strategy for all 5 types of models on both the Google and Backblaze datasets.
On the Alibaba dataset, the DDM concept drift retraining strategy reduces to training time by 50.9\% to 78.4\% for all five types of models compared with the periodical retraining strategy, while the STEPD concept drift retraining strategy has a similar retraining time compared with the periodical retraining strategy (the STEPD retrained on every new training period on the Alibaba dataset).
The PERM concept drift guided retraining strategy reduces the training time (i.e., 35.9\% to 71.2\%) on the Google dataset.
On the Backblaze dataset, the PERM concept drift guided retraining strategy reduces the training time for only the LR model (55.2\% reduction) but increases the training time for the other 4 types of models (12.7\%, 49.8\%, 42.2\%, and 14.5\% increase of training time for the CART, GBDT, NN, and RF model, respectively).
On the Alibaba dataset, the PERM concept drift guided retraining strategy shows an increase in the training time on all five types of models ranging from 13.9\% to 23.8\%. 
The PERM concept drift guided retraining strategy involves additional model training for each time period to detect potential concept drift and the situation, slowing the training performance, especially on the Alibaba and Backblaze datasets, as their retrain frequency is higher.
On the other hand, the time-based ensembles have an even shorter training time for some models while increasing the training time for others compared with periodical retraining strategies.
For example, the SEA ensemble trains faster on all 5 models on the Google dataset, 3 out of 5 models on the Alibaba dataset (on CART, RF, and GBDT), and 2 out of 5 models on the Backblaze dataset (i.e., LR and CART) than the fastest concept drift guided retraining strategy while being even slower than the periodical retraining strategy for 2 out of 5 models (i.e., NN and GBDT) on the Backblaze dataset.
Time-based ensemble strategies must construct and evaluate base models for each time period. 
As the Backblaze dataset is enormous in volume (even a single time period holds hundreds of thousands of data samples), constructing and evaluating these base models can be very slow, which may explain the slow training speed of the time-based ensemble strategies on the Backblaze dataset for complex models (e.g., NN and GBDT).
The online learning strategies can also take a shorter training time than other model update strategies. 
For example, AUE online learning strategy, the best-performing strategy on the Backblaze dataset, takes only 1,454 seconds to train and update the model; in comparison, the second-best-performing models on the Backblaze dataset (i.e., Retrain/NN and STEPD/NN), take 8,720 and 5,738 seconds, respectively.

\textbf{While time-based ensemble strategies and online learning strategies can save model training time, they take significantly longer time in model testing than other mode update strategies}. 
Time-based ensemble strategies and online learning strategies can usually save training time and be used in situations where retraining the model from scratch is too expensive~\cite{domingos2000mining, wang2003mining}.
However, as shown in Table~\ref{tab:effort_table}, the online learning strategies and time-based ensemble strategies can take much longer testing time than other model update strategies. 
For example, on the Backblaze dataset, the online learning strategies take an average of 3,802 to 62,810 seconds to test the model on the testing time periods, and the time-based ensemble strategies take up to 4,132 seconds for testing.
In comparison, other strategies that only use base models (i.e., stationary and retrained models) take less than 358 seconds to test the model on the testing time periods.
For the same model (e.g., RF), the time-based ensemble strategies usually take tens of times the testing time of strategies that only use base models.
In particular, while other strategies take shorter testing time than the corresponding training time, the online learning strategies take even longer testing time than training time. 
For example, the AUE strategy takes 43 times longer to test (62,810 seconds) than to train (1,454 seconds) on the Backblaze dataset.
Online learning models preserve the ability to modify the internal structure, thus increasing the time needed for inference~\cite{tan2011fast}.
In conclusion, time-based ensemble strategies and online learning strategies save training time by increasing the complexity of the models at the cost of significantly increasing the testing time.
As operation data usually arrive in very large batches (i.e., hundreds of thousands of data instances) and the delay in mitigating incidents would cause tremendous cost, such slow-in-testing strategies may hinder the ability of AIOps solutions to provide timely predictions~\cite{xu2018improving, lou2013software}.
Emerging new research strategies (e.g., online learning and time-based ensemble) are still not mature enough to be used in practice, although they are showing better performance compared to traditional retraining strategies.

\subsection{Stability} 
\begin{figure}[!ht]
    \centering
    \subfloat[Google]{\label{fig:cv_sk_google}{
        \includegraphics[width=0.95\textwidth]{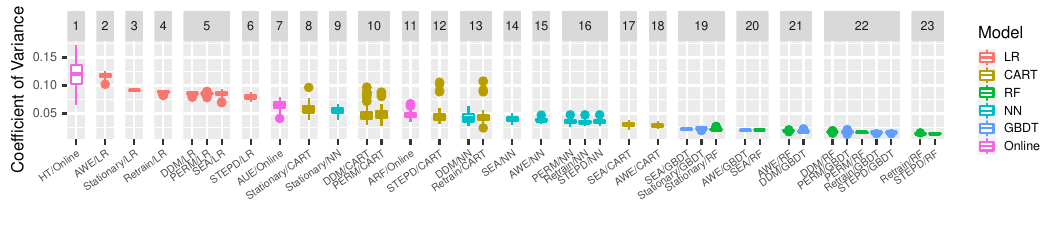}\hfill
    }}\hfill
    \subfloat[Backblaze]{\label{fig:cv_sk_backblaze}{
        \includegraphics[width=0.95\textwidth]{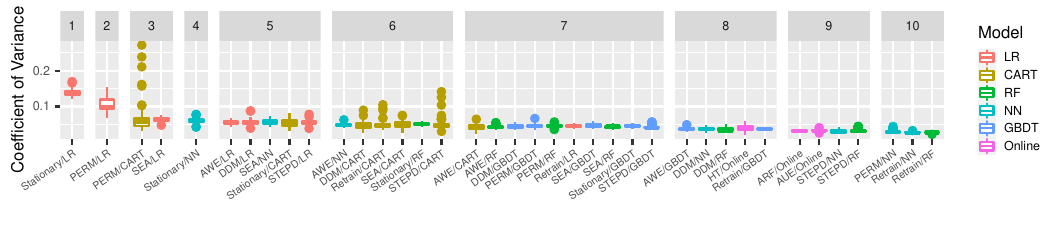}\hfill
    }}\hfill
    \subfloat[Alibaba]{\label{fig:cv_sk_alibaba}{
        \includegraphics[width=0.95\textwidth]{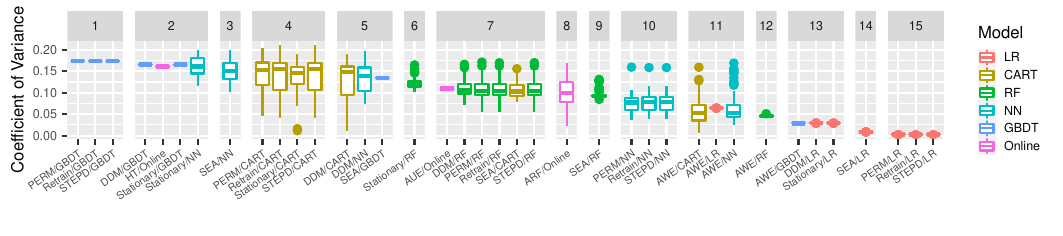}\hfill
    }}
    \caption{Scott-Knott test results of the coefficient of variance of the AUC performance of different model update strategies.}
    \label{fig:cv_sk}
\end{figure}

Table~\ref{tab:stability_table} shows, for each combination of model update strategy and model choice, the coefficient of variance (CV) of the AUC over the testing periods.
Figure~\ref{fig:cv_sk} shows our Scott-Knott test results of ranking the combinations according to their CV values.

\textbf{Active model update strategies (i.e., periodical retraining, concept drift guided retraining, and time-based ensemble strategies) achieve more stable performance than stationary model}.
As shown in Table~\ref{tab:stability_table}, the studied model update strategies, including the periodical retraining strategy, the concept drift guided retraining strategies, and the time-based ensemble strategies, all show smaller performance variance than the stationary models for most of the cases.
In particular, the periodical retraining strategy usually shows the stablest results (i.e., with the smallest CV values) among all the model update strategies.
As shown in our Scott-Knott test results ((Figure~\ref{fig:cv_sk})), the periodical retraining strategy shows the stablest results for 4 out of 5 base models (except for LR) on the Google dataset, 4 out of the 5 base models (except for CART) on the Backblaze dataset, and 3 out of the 5 base models (except for GBDT and CART) on the Alibaba dataset.
Since the stationary model may suffer from performance degradation caused by concept drift while actively maintaining models could mitigate the concept drift problem and sustain model performance, it is reasonable that the model update strategies achieve higher stability in terms of AUC performance.`'
In general, more frequent model updating is needed to achieve more stable model performance.

\begin{table}[!ht]
    \centering 
    \caption{The stability evaluation of different model maintenance strategies in terms of the coefficient of variance (CV) of AUC on each testing period.}
    \label{tab:stability_table}
    \resizebox{\textwidth}{!}{
    \begin{threeparttable}
    \begin{tabular}{*{3}{c}*{10}{l}}
        \toprule
        \multirow{3}{*}{Item}& \multirow{3}{*}{Dataset}& \multirow{3}{*}{Model} & \multicolumn{10}{c}{Strategy}\\
        \cline{4-13}
        & & & \multirow{2}{*}{Stationary}& \multirow{2}{*}{Retrain}& \multicolumn{3}{c}{Detection}& \multicolumn{2}{c}{Emsemble}& \multicolumn{3}{c}{Online}\\
        \cline{6-13}
        & & & & & DDM& PERM& STEPD& SEA& AWE& HT& ARF& AUE\\
        \midrule
        \multirow{15}{*}{CV~\tnote{1}}& \multirow{5}{*}{Google}& 
            LR & 0.09 (0.0\%) & 0.09 (-3.1\%) & 0.09 (-6.3\%) & 0.09 (-6.5\%) & 0.08 (-12.9\%) & 0.09 (-6.5\%) & 0.12 (28.6\%) &\multirow{5}{*}{0.12~\tnote{2}}& \multirow{5}{*}{0.05}& \multirow{5}{*}{0.06}\\
        & & CART & 0.06 (0.0\%) & 0.04 (-24.1\%) & 0.05 (-13.9\%) & 0.05 (-14.7\%) & 0.05 (-21.4\%) & 0.03 (-48.5\%) & 0.03 (-52.0\%) \\ 
        & & RF & 0.02 (0.0\%) & 0.01 (-37.5\%) & 0.02 (-23.3\%) & 0.02 (-25.4\%) & 0.01 (-37.1\%) & 0.02 (-5.6\%) & 0.02 (-12.0\%) \\ 
        & & NN & 0.05 (0.0\%) & 0.03 (-37.1\%) & 0.04 (-22.4\%) & 0.04 (-35.6\%) & 0.04 (-36.1\%) & 0.04 (-28.1\%) & 0.04 (-31.1\%) \\ 
        & & GBDT & 0.02 (0.0\%) & 0.02 (-30.2\%) & 0.02 (-22.7\%) & 0.02 (-27.2\%) & 0.02 (-29.9\%) & 0.02 (-4.6\%) & 0.02 (-12.8\%) \\
        \cline{2-13}
        & \multirow{5}{*}{Backblaze}&
            LR & 0.14 (0.0\%) & 0.05 (-66.8\%) & 0.06 (-59.4\%) & 0.10 (-25.2\%) & 0.05 (-60.6\%) & 0.06 (-54.7\%) & 0.05 (-60.4\%) &\multirow{5}{*}{0.04}& \multirow{5}{*}{0.03}& \multirow{5}{*}{0.03}\\
        & & CART & 0.05 (0.0\%) & 0.05 (-11.4\%) & 0.05 (-13.6\%) & 0.07 (20.4\%) & 0.05 (-9.3\%) & 0.05 (-6.6\%) & 0.04 (-21.0\%) \\ 
        & & RF & 0.05 (0.0\%) & 0.03 (-45.5\%) & 0.03 (-32.1\%) & 0.05 (-8.1\%) & 0.03 (-36.3\%) & 0.04 (-13.0\%) & 0.04 (-14.5\%) \\ 
        & & NN & 0.06 (0.0\%) & 0.03 (-55.7\%) & 0.04 (-38.0\%) & 0.03 (-51.3\%) & 0.03 (-48.2\%) & 0.06 (-6.4\%) & 0.05 (-18.9\%) \\ 
        & & GBDT & 0.05 (0.0\%) & 0.04 (-18.5\%) & 0.04 (-1.7\%) & 0.05 (1.1\%) & 0.04 (-7.7\%) & 0.05 (2.3\%) & 0.04 (-18.5\%) \\ 
        \cline{2-13}
        & \multirow{5}{*}{Alibaba}&
            LR & 0.03 (0.0\%) & 0.00 (-91.5\%) & 0.03 (0.0\%) & 0.00 (-91.5\%) & 0.00 (-91.5\%) & 0.01 (-69.8\%) & 0.06 (120.6\%) &\multirow{5}{*}{0.16}& \multirow{5}{*}{0.10}& \multirow{5}{*}{0.11}\\
        & & CART & 0.14 (0.0\%) & 0.14 (1.5\%) & 0.13 (-4.1\%) & 0.14 (1.1\%) & 0.14 (1.5\%) & 0.11 (-22.8\%) & 0.06 (-58.1\%) \\ 
        & & RF & 0.12 (0.0\%) & 0.11 (-11.2\%) & 0.11 (-10.3\%) & 0.11 (-11.1\%) & 0.11 (-11.2\%) & 0.10 (-21.6\%) & 0.05 (-62.4\%) \\ 
        & & NN & 0.16 (0.0\%) & 0.08 (-52.9\%) & 0.13 (-17.3\%) & 0.08 (-53.6\%) & 0.08 (-52.9\%) & 0.15 (-7.3\%) & 0.06 (-61.6\%) \\ 
        & & GBDT & 0.17 (0.0\%) & 0.17 (5.0\%) & 0.17 (0.3\%) & 0.17 (5.0\%) & 0.17 (5.0\%) & 0.13 (-18.4\%) & 0.03 (-82.7\%) \\ 
        \bottomrule
    \end{tabular}
    \begin{tablenotes}
    \item[1] The number in brackets shows the relative increase of CV compared with corresponding stationary models.
    \item[2] We skip the calculation of performance improvement for online models as they do not have corresponding stationary models.
    \end{tablenotes}
    \end{threeparttable}
    }
\end{table}

\textbf{The stability of different update strategies and model types is highly correlated with its performance: a better performing one is usually more stable}.
As shown in Table~\ref{tab:stability_table}, on the Google dataset, the RF model, the GBDT model, and the CART model with the time-based ensemble strategies are among the stablest strategies.
These strategies are also the best-performing ones (with the highest AUC values) on the Google dataset (as shown in Figure~\ref{fig:auc_sk}).
On the Backblaze dataset, the online learning strategies (AUE, ARF, and HT) are among the most stable strategies and are among the best-performing ones on the same dataset.
Similarly, the best-performing ensemble strategies are among the stablest strategies.
In fact, we measure the Spearman correlation between the stability and performance (AUC) of the combinations of model update strategies and model choices, and we observe a high correlation value of -0.93 (i.e., a higher AUC is correlated with a smaller CV).

\section{Discussion}
\label{sec:discussion}

\textbf{Model retraining may not always lead to performance improvement of AIOps models in the immediate future. However, adopting a higher model retraining frequency strategy can always lead to better performance in the long term}.
In order to track the performance impact of model retraining, we compare the periodical retraining approach, the three concept drift guided retraining approaches, and a \emph{hypothesized optimal} approach in terms of their retrain frequency and performance improvement over the stationary approach.
The hypothesized optimal approach is based on the intuition that a model needs to be retrained when retraining the model statistically significantly improves the model performance on the samples in the following time period.
Specifically, when the data from a new time period arrives, the optimal approach trains a new model with the latest data and compares the performance of the new model with the performance of the previously-trained model on the data samples from the following time period (i.e., a future time period). 
Upon detecting a statistically significant difference between the two models' performance, we deem that concept drift occurred and replace the previous model with the new one. Otherwise, we retain the old, previously trained model.
In particular, we train and test the models $100$ times and use a Mann-Whitney U test (p $\le$ 0.05) to quantify the statistical significance of the performance difference between model performance (i.e., AUC) before and after an update.
In a practical setting, we would not have the knowledge of how the model performs on future data. Hence, it is a \textit{hypothetical optimal} approach in which we leverage historical data to simulate and assess the evaluation criteria of model updates in different strategies. 

Table~\ref{tab:detection_google} and Table~\ref{tab:detection_disk} show the impact of the model retraining approaches on each testing time period for the retraining approaches on the Google and Backblaze datasets, respectively. 
We exclude the Alibaba dataset from the discussion because only 3 retraining and testing periods are available for it.
As shown in the results, a higher number of retrains generally produces higher performance improvement.
In some cases, the ``optimal'' approach achieves a better performance than the periodical retraining approach (e.g., for the CART model on the Google dataset); however, in other cases, the periodical retraining approach performs better (e.g., for the NN model applied on the Backblaze dataset). 
While the ``optimal'' approach acts as a ``prophet'' and only retrains a model when it could improve the performance, it evaluates the performance impact on the following time period instead of on all future time periods, thus it may not lead to the best overall performance in all cases.
In addition, even if retraining a model could negatively impact its performance in the immediate next time period, the model retraining decision could still benefit the model performance in the long term. 
For example, the PERM concept guided retraining approach on the LR model suggests one more retraining (i.e., for the 18th time period) than the DDM concept guided retraining approach on the Google dataset. Although the extra retrain decreases the model performance in the immediate subsequent time period, the PERM approach still results in a 1.1\% more overall performance increase.

\begin{table}[!ht]
    \centering
    \caption{The detection of concept drift in each time period by different retraining strategies on the Google cluster trace dataset. A symbol indicates that an occurrence of concept drift is detected in the corresponding time period by the corresponding strategy on over half of the experiment rounds. An $\uparrow$ means the AUC performance significantly increased, a $-$ means no significant difference in AUC performance, and a $\downarrow$ means the AUC performance significantly decreased, compared to not updating the model using the Mann–Whitney U Test.}
    \label{tab:detection_google}
    \begin{threeparttable}
        \begin{tabular}{*{16}c}
            \toprule
            \multirow{2}{*}{Model}& \multirow{2}{*}{Strategy}& \multicolumn{13}{c}{Testing period}& \multirow{2}{*}{\specialcell{Performance\\Improvement}}\\
            \cline{3-15}
            & &16& 17& 18& 19& 20& 21& 22& 23& 24& 25& 26& 27& 28\\
            \midrule
            \multirow{5}{*}{LR}
            & Retrain& $\downarrow$& $\downarrow$& $\downarrow$& $\downarrow$& $\uparrow$& $\uparrow$& $\uparrow$& $\downarrow$& $\downarrow$& $\downarrow$& $\downarrow$& $\uparrow$& $\downarrow$& 10.3\% \\
            & DDM& & & & & $\uparrow$& & & $\downarrow$& & & & & & 5.6\% \\
            & PERM& & & $\downarrow$& & $\uparrow$& & & $\downarrow$& & & & & & 6.7\% \\
            & STEPD& $\downarrow$& & & $\downarrow$& $\uparrow$& $\uparrow$& $\uparrow$& $\downarrow$& $\downarrow$& & & & $\downarrow$& 10.0\% \\
            & Optimal& & $\uparrow$& & & & & $\uparrow$& $\uparrow$& & $\uparrow$& $\uparrow$& & $\uparrow$& 8.9\% \\
            \midrule
            \multirow{5}{*}{CART}
            & Retrain& $\uparrow$& $\uparrow$& $\uparrow$& $\uparrow$& $-$& $\uparrow$& $-$& $-$& $\uparrow$& $\uparrow$& $\uparrow$& $\uparrow$& $\uparrow$& 2.0\% \\
            & DDM& & & $\uparrow$& & $\uparrow$& & & $-$& & & & & & 1.0\% \\
            & PERM& $\uparrow$& $\uparrow$& $\uparrow$& & $\uparrow$& & & $-$& & & & $\uparrow$& & 1.0\% \\
            & STEPD& $\uparrow$& $\uparrow$& $\uparrow$& $\uparrow$& $-$& & $\uparrow$& $-$& & $\uparrow$& $\uparrow$& $\uparrow$& $\uparrow$& 1.9\% \\
            & Optimal& $\uparrow$& $\uparrow$& $\uparrow$& $\uparrow$& & $\uparrow$& & & $\uparrow$& $\uparrow$& $\uparrow$& $\uparrow$& $\uparrow$& 3.2\% \\
            \midrule
            \multirow{5}{*}{RF}
            & Retrain& $\uparrow$& $\uparrow$& $\uparrow$& $\uparrow$& $\uparrow$& $\uparrow$& $-$& $\uparrow$& $\uparrow$& $\uparrow$& $\uparrow$& $\uparrow$& $\uparrow$& 1.5\% \\
            & DDM& & $\uparrow$& & & $\uparrow$& & & $\uparrow$& & & & & & 1.0\% \\
            & PERM& $\uparrow$& $\uparrow$& $\uparrow$& & $\uparrow$& & & $\uparrow$& & & & $\uparrow$& & 1.1\% \\
            & STEPD& $\uparrow$& $\uparrow$& & $\uparrow$& $\uparrow$& & & $\uparrow$& $\uparrow$& $\uparrow$& $\uparrow$& $\uparrow$& $\uparrow$& 1.5\% \\
            & Optimal& $\uparrow$& $\uparrow$& $\uparrow$& $\uparrow$& $\uparrow$& $\uparrow$& & $\uparrow$& $\uparrow$& $\uparrow$& $\uparrow$& $\uparrow$& $\uparrow$& 1.3\% \\
            \midrule
            \multirow{5}{*}{NN}
            & Retrain& $\uparrow$& $\uparrow$& $-$& $\uparrow$& $-$& $\downarrow$& $\uparrow$& $\uparrow$& $\uparrow$& $-$& $\uparrow$& $\uparrow$& $\uparrow$& 4.6\% \\
            & DDM& & & & & & & & & & & & $\uparrow$& & 2.4\% \\
            & PERM& $\uparrow$& $\uparrow$& $-$& & $\uparrow$& & & $\uparrow$& & & & $\uparrow$& & 4.1\% \\
            & STEPD& $\uparrow$& $\uparrow$& $-$& $\uparrow$& $-$& $\downarrow$& $\uparrow$& $\uparrow$& & & $\uparrow$& $\uparrow$& $\uparrow$& 4.4\% \\
            & Optimal& $\uparrow$& $\uparrow$& & $\uparrow$& & $\uparrow$& $\uparrow$& $\uparrow$& $\uparrow$& & $\uparrow$& $\uparrow$& $\uparrow$& 4.1\% \\
            \midrule
            \multirow{5}{*}{GBDT}
            & Retrain& $-$& $\uparrow$& $\downarrow$& $\uparrow$& $\uparrow$& $\uparrow$& $\uparrow$& $\uparrow$& $\uparrow$& $\uparrow$& $\uparrow$& $\uparrow$& $\uparrow$& 1.5\% \\
            & DDM& & $\uparrow$& & & $\uparrow$& & & $\uparrow$& & & & & & 1.0\% \\
            & PERM& $-$& $\uparrow$& $\downarrow$& & $\uparrow$& & & $\uparrow$& & $\uparrow$& & & & 1.1\% \\
            & STEPD& $-$& $\uparrow$& $\downarrow$& $\uparrow$& $\uparrow$& $\uparrow$& $\uparrow$& $\uparrow$& $\uparrow$& $\uparrow$& $\uparrow$& $\uparrow$& $\uparrow$& 1.5\% \\
            & Optimal& & $\uparrow$& & $\uparrow$& $\uparrow$& $\uparrow$& $\uparrow$& $\uparrow$& $\uparrow$& $\uparrow$& $\uparrow$& $\uparrow$& $\uparrow$& 1.3\% \\
            \bottomrule
        \end{tabular}
    \begin{tablenotes}
    \item[1] The performance improvement indicates the relative improvement of the overall AUC performance against the corresponding stationary models.
    \end{tablenotes}
    \end{threeparttable}
\end{table}

\begin{table}[!ht]
    \centering
    \caption{The detection of concept drift in each time period by different retraining strategies on the Backblaze disk stats dataset. A symbol indicates that an occurrence of concept drift is detected in the corresponding time period by the corresponding strategy on over half of the experiment rounds. An $\uparrow$ means the AUC performance significantly increased, a $-$ means no significant difference in AUC performance, and a $\downarrow$ means the AUC performance significantly decreased, compared to not updating the model using the Mann–Whitney U Test.}
    \label{tab:detection_disk}
    \resizebox{\textwidth}{!}{
    \begin{threeparttable}
        \begin{tabular}{*{20}c}
            \toprule
            \multirow{2}{*}{Model}& \multirow{2}{*}{Strategy}& \multicolumn{17}{c}{Testing period}& \multirow{2}{*}{\specialcell{Performance\\Improvement}}\\
            \cline{3-19}
            & &20& 21& 22& 23& 24& 25& 26& 27& 28& 29& 30& 31& 32& 33& 34& 35& 36\\
            \midrule
            \multirow{5}{*}{LR}
            & Retrain& $\downarrow$& $\uparrow$& $\uparrow$& $\downarrow$& $\uparrow$& $\uparrow$& $\uparrow$& $\downarrow$& $\uparrow$& $\uparrow$& $\uparrow$& $\uparrow$& $\downarrow$& $-$& $\downarrow$& $\uparrow$& $\uparrow$& 25.4\% \\
            & DDM& & & & & $\uparrow$& & & & & & $\uparrow$& & & & $\uparrow$& & & 23.9\% \\
            & PERM& & & & & $\uparrow$& & & & & & & & & & & & & 11.0\% \\
            & STEPD& $\downarrow$& & & $\uparrow$& $\uparrow$& $-$& $-$& & & $\uparrow$& $\uparrow$& $\uparrow$& $\downarrow$& $-$& $\downarrow$& $\uparrow$& & 24.5\% \\
            & Optimal& & $\uparrow$& $\uparrow$& & $\uparrow$& $\uparrow$& $\uparrow$& & $\uparrow$& $\uparrow$& $\uparrow$& $\uparrow$& & $\uparrow$& & $\uparrow$& $\uparrow$ & 23.1\% \\
            \midrule
            \multirow{5}{*}{CART}
            & Retrain& $\uparrow$& $\uparrow$& $\uparrow$& $\uparrow$& $-$& $\uparrow$& $\uparrow$& $\uparrow$& $\uparrow$& $\uparrow$& $\uparrow$& $\downarrow$& $\uparrow$& $\downarrow$& $\uparrow$& $-$& $\uparrow$& 7.9\% \\
            & DDM& & & & & & & $\uparrow$& & & & & & & & & & & 7.0\% \\
            & PERM& & & & & & & & & & & & & & & & & & 4.9\% \\
            & STEPD& $\uparrow$& $\uparrow$& $\uparrow$& $\uparrow$& $-$& $\uparrow$& $\uparrow$& $\uparrow$& $\uparrow$& $\uparrow$& $\uparrow$& $\downarrow$& $\uparrow$& $\downarrow$& $\uparrow$& $-$& $\uparrow$& 8.1\% \\
            & Optimal& $\uparrow$& $\uparrow$& $\uparrow$& $\uparrow$& & $\uparrow$& $\uparrow$& $\uparrow$& $\uparrow$& $\uparrow$& $\uparrow$& & $\uparrow$& & & & & 9.1\% \\
            \midrule
            \multirow{5}{*}{RF}
            & Retrain& $\uparrow$& $\uparrow$& $\uparrow$& $\uparrow$& $\downarrow$& $\downarrow$& $\downarrow$& $\uparrow$& $\uparrow$& $\downarrow$& $\uparrow$& $\uparrow$& $-$& $\uparrow$& $\uparrow$& $-$& $\uparrow$& 3.8\% \\
            & DDM& & & & & & & & & $\uparrow$& & & & & & & & & 4.0\% \\
            & PERM& & & & & $\downarrow$& & & & & & & & & & & & & -0.3\% \\
            & STEPD& & & & $\uparrow$& $\downarrow$& & $\uparrow$& & $\uparrow$& $\downarrow$& $\uparrow$& $\uparrow$& $-$& $\uparrow$& $\uparrow$& $-$& $\uparrow$& 3.8\% \\
            & Optimal& $\uparrow$& $\uparrow$& $\uparrow$& $\uparrow$& & & & $\uparrow$& $\uparrow$& & $\uparrow$& $\uparrow$& & $\uparrow$& $\uparrow$& & $\uparrow$& 4.7\% \\
            \midrule
            \multirow{5}{*}{NN}
            & Retrain& $-$& $\uparrow$& $\uparrow$& $\uparrow$& $\downarrow$& $\downarrow$& $\uparrow$& $\uparrow$& $\uparrow$& $\uparrow$& $\uparrow$& $\uparrow$& $\downarrow$& $\uparrow$& $\uparrow$& $\uparrow$& $\uparrow$& 7.8\% \\
            & DDM& & & & & & & $\uparrow$& & & & & $\uparrow$& & & $\uparrow$& & & 6.9\% \\
            & PERM& & & & & $\downarrow$& & $\uparrow$& $\uparrow$& $\uparrow$& & $\uparrow$& & $\uparrow$& & & & $\uparrow$& 7.0\% \\
            & STEPD& & & $\uparrow$& & $\downarrow$& & $\uparrow$& & & $\uparrow$& $\uparrow$& $\uparrow$& $\downarrow$& & $\uparrow$& $\uparrow$& $\uparrow$& 7.6\% \\
            & Optimal& & & $\uparrow$& $\uparrow$& & & $\uparrow$& $\uparrow$& $\uparrow$& $\uparrow$& $\uparrow$& $\uparrow$& & $\uparrow$& $\uparrow$& $\uparrow$& $\uparrow$& 7.0\% \\
            \midrule
            \multirow{5}{*}{GBDT}
            & Retrain& $\uparrow$& $\downarrow$& $\uparrow$& $\downarrow$& $\uparrow$& $\downarrow$& $\uparrow$& $\uparrow$& $\uparrow$& $\uparrow$& $\uparrow$& $-$& $\uparrow$& $\downarrow$& $\uparrow$& $\downarrow$& $\uparrow$& 5.0\% \\
            & DDM& & & & & & & $\uparrow$& & & & & & & & $\uparrow$& & $\uparrow$& 4.1\% \\
            & PERM& & & & & & & & & & & & $\uparrow$& & & & & & 3.8\% \\
            & STEPD& $\uparrow$& & $\uparrow$& & $\downarrow$& & $\uparrow$& & $\uparrow$& $\uparrow$& $\uparrow$& $-$& $\uparrow$& & $\uparrow$& $\downarrow$& $\uparrow$& 4.8\% \\
            & Optimal& $\uparrow$& & $\uparrow$& & & $\uparrow$& $\uparrow$& $\uparrow$& $\uparrow$& $\uparrow$& $\uparrow$& & $\uparrow$& & $\uparrow$& & $\uparrow$& 5.9\% \\
            \bottomrule
        \end{tabular}%
    \begin{tablenotes}
    \item[1] The performance improvement indicates the relative improvement of the overall AUC performance against the corresponding stationary models.
    \end{tablenotes}
    \end{threeparttable}
        }
\end{table}

\begin{table}[!ht]
    \centering 
    \caption{The comparison of different model maintenance strategies along each evaluation dimension.}
    \label{tab:comparison_table}
    \resizebox{\textwidth}{!}{
    \begin{threeparttable}
    \begin{tabular}{*{11}{c}}
        \toprule 
        \multirow{3}{*}{Evaluation dimension}& \multicolumn{10}{c}{Strategy~\tnote{1}}\\
        \cline{2-11}
        & \multirow{2}{*}{Stationary}& \multirow{2}{*}{Retrain}& \multicolumn{3}{c}{Detection}& \multicolumn{2}{c}{Emsemble}& \multicolumn{3}{c}{Online}\\
        \cline{4-11}
        & & & DDM& PERM& STEPD& SEA& AWE& HT& ARF& AUE\\
        \midrule
        Performance& $\times$& \checkmark& & $\circ$& \checkmark & $\circ$& \checkmark& & $\circ$& \\
        Cost-effectiveness & & $\times$& \checkmark& \checkmark& $\circ$& $\times$& $\times$& $\times$& $\times$& $\times$\\
        Training time & \checkmark& $\times$& & \checkmark & & \checkmark& \checkmark & & & \\
        Testing time & & & & & & $\times$& $\times$& $\times$& $\times$& $\times$\\
        Stability & $\times$& \checkmark& \checkmark& $\circ$ & \checkmark& & \checkmark \\
        \bottomrule
    \end{tabular}
    \begin{tablenotes}
    \item[1] A checkmark (\checkmark) indicates a recommended strategy based on the performance for the majority of scenarios in the corresponding evaluation dimension, a circle ($\circ$) indicates less recommended strategies with medium performance, while a cross mark ($\times$) indicates a not recommended strategies due to performance in certain evaluation dimensions. 
    \end{tablenotes}
    \end{threeparttable}
    }
\end{table}

\textbf{Practitioners should be cautious while deciding when and how to maintain their AIOps models, depending on their performance requirements, resources, and the characteristics of data.}
As retraining AIOps models come with benefits (e.g., performance improvement and higher stability) and costs (e.g., computing cost for maintenance), practitioners should be cautious when adapting model update strategies. 
We summarize the performance of different model update strategies along each evaluation dimension (i.e., AUC performance, EC ratio, computing time, and stability) and suggest one or several general picks for each dimension in Table~\ref{tab:comparison_table}.
We also advise that different approaches have their own advantages on specific evaluation dimensions, and there is no clear winner on all of the evaluation dimensions.
Hence, practitioners should consider the requirements and constraints of their application scenario to trade-off between the model performance, training effort, stability, and other dimensions.
For example, in our case study on the Backblaze disk stat dataset, the new data feed in at a steady and relatively low pace (i.e., daily snapshot for each machine) and does not require immediate responses. 
Hence, practitioners could deploy the maintenance-heavy time-based ensemble approaches and schedule the prediction and model maintenance daily to increase prediction performance while bringing little maintenance overhead.
On the other hand, our case study on the Google and Alibaba cluster trace datasets indicates a need for high inference speed and minimum downtime. The periodical retraining approach is then more suitable than the time-based ensemble and online learning approaches, which have a slower inference speed or need constant model updates.

\textbf{Practitioners could consider advanced model update strategies (e.g., concept-drift guided retraining or time-based ensembles) as alternatives to periodical model retraining}.
The industry usually applies periodical retraining approaches to maintain the model performance. 
However, our case study shows that advanced model update strategies can reach similar or even higher performance while bringing additional advantages.
For example, both the SEA and AWE time-based ensemble approaches achieve higher AUC performance on the GBDT model, with an increase of 8.4\% and 13.2\% over the periodical retraining approach on the Alibaba dataset, as shown in Table~\ref{tab:performance_table}.
On the other hand, although not outperforming the periodical retraining approach, concept drift guided retraining approaches can reach statistically comparable performance to the periodical retraining approaches.
Besides, the concept drift guided retraining approaches can significantly reduce the frequency of model retraining compared with periodical retraining (e.g., a reduction between 76.9\% and 84.6\% on the Google dataset and 64.7\% to 82.3\% on the Backblaze dataset when using the DDM concept drift guided retraining approach).
In addition, except that the PERM concept drift guided retraining approach requires additional model retraining to operate, the DDM and STEPD concept drift guided retraining approaches only add insignificant time for concept drift detection compared with the model training cost.
Although the time-based ensemble approaches require the same amount of retraining as the periodical retraining approach, they significantly reduce the retraining cost (e.g., 76.4\% to 96.7\% reduction in training time on the Google dataset when using the SEA ensemble approach) since only one small base classifier is trained when the newest data is added.

\textbf{Future works are needed for improving the efficiency of the concept drift guided retraining and time-based ensemble approaches}.
We investigated why some concept drift guided retraining approaches may require longer training time than others (e.g., the PERM retraining approach even has longer training time than the periodical retraining approach on 4 out of 5 models on the Backblaze data), and found that it accounts for both the time spent on detecting concept drift and the frequency of retraining. 
When a new time period becomes available, extra model training and evaluation are performed to detect concept drift.
For each new time period, the PERM approach performs $100$ random splits of the samples in the period and builds a model for each split to detect concept drift. 
The DDM and STEPD approaches do not involve extra model training and only require extra model testing on the new time period to detect concept drift; hence they are faster for detecting concept drift.
As shown in Table~\ref{tab:ec_table}, the STEPD approach reports concept drift more often than the other two concept drift detection methods, which explains why it is slower in certain scenarios.
A similar circumstance happens in the time-based ensemble approaches.
For example, both time-based ensemble approaches are slower in training than the periodical retraining models on the NN and GBDT models on the Backblaze dataset, as shown in Table~\ref{tab:effort_table}.
The reason is that the ensemble approaches need extra time to build a new classifier on each new time period and evaluate all base classifiers in the ensemble on the data in the new time period to determine whether and how to replace base classifiers (for both SEA and AWE) and assign weights to the base classifiers (for AWE).
Therefore, we suggest that future research could develop methods that can efficiently detect concept drift in operation data (usually massive in volume) and more efficient time-based ensemble algorithms.

\textbf{The efficiency of model testing may be equally important as the efficiency of model training for AIOps solutions where the data arrive at high speed and fast model inferences are required}.
As shown in our experiment results (Table~\ref{tab:effort_table}), the time-based ensemble approaches and online learning approaches tend to consume a large amount of time in testing.
When a new time period becomes available, the time-based ensemble approaches only train a base classifier on the data from the new time period. 
Therefore, using time-based ensemble approaches to update AIOps models can be more efficient than retraining a single model using all the training data.
Prior work also reports that time-based ensemble approaches can lead to shorter training time~\cite{wang2003mining}.
However, during model testing, such ensemble approaches need to evaluate all the base classifiers on the testing data, which can lead to a very slow testing speed.
Similarly, online learning approaches can save training time but at the cost of longer testing time. 
Prior work notes that the need for modifying the model trees (i.e., to allow online learning) causes HT's runtime to be four to six times slower than the tree model that does not require structural modifications~\cite{tan2011fast}.
Hence, while online learning approaches are able to be updated on the fly by preserving the ability to modify the internal model structure, such ability can significantly sacrifice model testing time.
As operation data usually arrives at high speed and volume, applying time-based ensemble or online learning approaches may not be ideal when the application requires applying the model on a large volume of samples simultaneously.

\textbf{Data pre-processing of operation data}.
Operation datasets are usually not presented in the form that general ML algorithms can readily consume.
For instance, the Google and Alibaba datasets are trace data of server workloads and machine monitoring, while the Backblaze dataset is in the form of daily snapshots of SMART diagnosis metrics.
All three datasets used in our case study need extensive pre-processing before turning into feature vectors that can be fed into ML models.
To understand the computational cost of data pre-processing, we measure the time consumption of data pre-processing for each of the three datasets.
The Google cluster trace dataset takes an average of 76.3 seconds on pre-processing for each of the one-day time periods while the Alibaba dataset takes only an average of 4.1 seconds on each of the one-week time periods.
Although the two datasets go through similar pre-processing steps, the discrepancy in pre-processing could be from the difference in how the job and monitoring information are recorded.
For example, the Google dataset records job and task status along certain intervals, requiring tracing down the status of each job and task through its life cycle, while the Alibaba dataset presents the job and task information in a single entry.
The Backblaze dataset also takes longer for pre-processing: we observe an average of 74.6 seconds for each of the one-month time periods.
The reasons for the prolonged pre-processing time could be both the sheer amount of data volume (i.e., an average of 114K samples per period for the Backblaze dataset vs. an average of 22K for the Google dataset) and the use of cumulative features that need calculation of the difference of value changes in the one-week windows.
Similar efforts may be needed for building AIOps solutions on other operation datasets, and we suggest practitioners keep in mind the pre-processing requirement and the associated extra time consumption in building their AIOps solutions.

\textbf{Generalizability of our results.} 
We discuss the generalizability of our results from two perspectives: 
1) which scenarios can our studied AIOps model update strategies and results apply to, and 2) the generalizability of our results to new datasets.

\begin{itemize}
    \item \textit{Scope of the application scenarios}. 
    In this work, we studied the evolution of the operation data distribution and the effects of different AIOps model update strategies in handling the evolution. 
    However, we did not consider other types of evolution in operation data, such as the changes in data format, or the addition or deletion of features that are caused by the evolution of the system under operation. 
    Our study also focuses on automated failure prediction, thus our results may not generalize to other types of AIOps tasks.
    Still, our studied model update strategies and results are applicable to other scenarios in automated failure prediction where the feature set and format formats do not change. 
    When the feature set or data formats are changed, human intervention is needed to perform feature engineering that considers the new features or new formats. 
    Nevertheless, even in such cases, our studied approaches and results are still applicable in the time periods when the feature sets and formats are stable. 
    We encourage future work to extend our work to consider other forms of data evolution.
    
    \item \textit{Generalizability to new datasets}.
    This work performs case studies on three public operation datasets (i.e., the Google cluster trace dataset, the Backblaze disk stats dataset, and the Alibaba GPU cluster trace dataset). 
    These are large-scale real-world datasets produced by production systems in industry from three different companies (i.e., Google, Backblaze, and Alibaba) regarding different aspects (i.e., CPU jobs, hard disks, and GPU jobs). 
    In order to improve the generalizability of our results, for each dataset, we consider five popular models (i.e., LR, CART, RF, NN, and GBDT) that are used in prior studies built on these datasets. 
    In fact, initially, our study was only performed on two of the three datasets (the Google cluster trace dataset and the Backblaze disk stats dataset). 
    Then, we added experiments on the third dataset (the Alibaba GPU cluster trace dataset) and observed that our initial results and conclusions still apply to the new dataset. 
    This observation demonstrates that our results and conclusions may generalize to more operation datasets. 
    Nevertheless, our work could benefit from evaluating the different AIOps model update strategies on more datasets, in particular, in a real-world application in an industry setting. 
    We encourage future work to extend our work by considering other datasets and models or performing an industry case study.
\end{itemize}

\section{Threats to Validity}
\label{sec:threats}

\subsection{External Validity}
A threat to the external validity of our findings concerns the generalizability of our results to other AIOps applications, datasets, and models. 
While we do not generalize our results, our study considers popular supervised models (i.e., LR, CART, RF, NN, and GBDT) on three datasets (i.e., the Google cluster trace dataset, the Backblaze disk stats dataset, and the Alibaba GPU cluster trace dataset) for two types of AIOps applications focusing on the automated prediction of failures.
Although these datasets were released from the industry (i.e., by Google, Backblaze, and Alibaba), our work could benefit from evaluating the different AIOps model update strategies in a real-world application in an industry setting. 
We encourage future work to extend our work by performing an industry case study.

Another external validity concerns the type of data evolution considered in this study.
We focus on studying the evolution of the data distribution of the features and their relationships rather than other kinds of evolution in operation data.
For example, we do not consider changes in data format, additions or deletions of features, or other possible forms of data evolution. 
We encourage future studies to extend our work to consider other forms of data evolution.

In addition, our evaluation approaches and criteria are all general and can be easily applicable to additional datasets or other model maintenance contexts.
Future work can replicate our study on more applications, datasets, and models.  

This work focuses on AIOps, which develops AI-based solutions to operationalize various software and system operations goals (e.g., job failure prediction or system anomaly detection).
Another highly correlated concept, MLOps~\cite{mlops}, instead focuses more on deploying and maintaining machine learning models in production reliably and efficiently.
After designing AIOps solutions, practitioners need MLOps tools to ensure the continuous delivery of machine learning solutions.
While our work on AIOps model updating can be considered an application area of MLOps, our work does not touch on automated tooling for the deployment and maintenance of general ML models.
Future research could consider incorporating MLOps tools for automating the delivery of AIOps solutions.

\subsection{Internal Validity}
A threat to the internal validity of our findings concerns the measurement of the training and testing time. 
Considering the time measurement is sensitive to many factors, we repeat our experiments $100$ times to mitigate noise in the measured training and testing time.

Our choice of the time period size could also be a threat to the validity of our results. 
Although the data could be partitioned into time periods in other sizes, we consider the natural time periods (e.g., days and months) such that practitioners can better leverage the tools to maintain their models (e.g., daily or monthly).

In this work, we consider two aspects (cost-effectiveness of model updating as well as model training and testing cost) to measure the updating cost of AIOps models, which focuses on the computational aspect of model maintenance.
However, there are other aspects of maintenance cost that are not considered in this work, such as the human efforts and resources taken to integrate, deploy, and monitor the AIOps model in the production system. 
In addition, the extra complexity of the system brought by the model update strategies themselves (e.g., the concept drift detection approaches) can introduce additional maintenance costs, such as the efforts needed for practitioners to learn the corresponding skills and maintain the implementation of the approaches. 
We plan to study these aspects in our future work.

\subsection{Construct Validity}
A threat to construct validity concerns our model training process. 
The configuration of the models may impact our results. In order to mitigate this impact, we use automated hyperparameter tuning (i.e., random search) to optimize the configuration of the hyperparameters, which is a widely used practice in the development of machine learning models. 
Nevertheless, different searches of the hyperparameters may lead to different results, and we cannot guarantee optimum hyperparameters. 
As our findings apply to all the studied datasets and models, we assume that using different hyper-parameters may not impact our findings. 

There may be other potential threats concerning our model training process. 
We use the same features as in prior works for the Google and Backblaze operation datasets and the same types of base models (except for the online learning approaches) appeared in the prior works~\cite{rosa2015predicting, elsayed2017learning, botezatu2016predicting, mahdisoltani2017proactive} (no prior work on predicting job failure is done on the Alibaba dataset) to reflect the training process in AIOps solutions.
Future work that evaluates our study by considering other modeling processes, e.g., using a different ML library or a different re-sampling approach, could benefit our study.

\section{Conclusion}
\label{sec:conclusion}
In this work, we study the model update strategies of supervised learning in the context of AIOps through a case study on three AIOps applications:
1) predicting job failures on the Google cluster trace data;
2) predicting disk failures on the Backblaze disk stats data;
3) predicting job failures on the Alibaba GPU cluster trace data.

We evaluate different model update strategies on our studied operation datasets along three dimensions: performance, model updating cost, and stability.
We observe that active model update strategies (i.e., periodical retraining, concept drift guided retraining, time-based ensemble, and online learning) all increase the model performance and stability compared to a stationary model.
In particular, concept drift guided retraining strategies can achieve similar model performance as periodical retraining while requiring fewer model updates.
Model updating strategies (i.e., time-based ensemble and online learning) could reduce the model training time but significantly increase the model testing time. 
Our findings suggest that AIOps practitioners should consider the evolving nature of their operation data and maintain their models with informed decisions (e.g., when there is concept drift) and cost-efficient update strategies.
Our results also indicate that no single strategy is best in all aspects (e.g., performance, model updating cost, and stability). 
Instead, practitioners should consider the context and product requirements (e.g., the urgency to deploy model updates and the characteristics of the dataset) to decide the most appropriate strategy.
We also suggest future research to investigate more efficient model update strategies that accommodate the unique characteristics of operation data (e.g., ultra-large-scale data and extremely imbalanced data distribution) and various deployment requirements (e.g., the urgency to push updates or the consideration of model maintenance efforts).

\bibliographystyle{ACM-Reference-Format}
\bibliography{aiops}

\end{document}